\documentclass[twocolumns]{sdl2}

\usepackage[T1]{fontenc}
\usepackage{lmodern}

\usepackage{colortbl}

\def\0{\mbox{\tiny $0$}}
\def\1{\mbox{\tiny $1$}}
\def\2{\mbox{\tiny $2$}}
\def\3{\mbox{\tiny $3$}}
\def\4{\mbox{\tiny $4$}}
\def\5{\mbox{\tiny $5$}}
\def\6{\mbox{\tiny $6$}}
\def\7{\mbox{\tiny $7$}}
\def\8{\mbox{\tiny $8$}}
\def\9{\mbox{\tiny $9$}}

\def\FigW{
\WideFigure{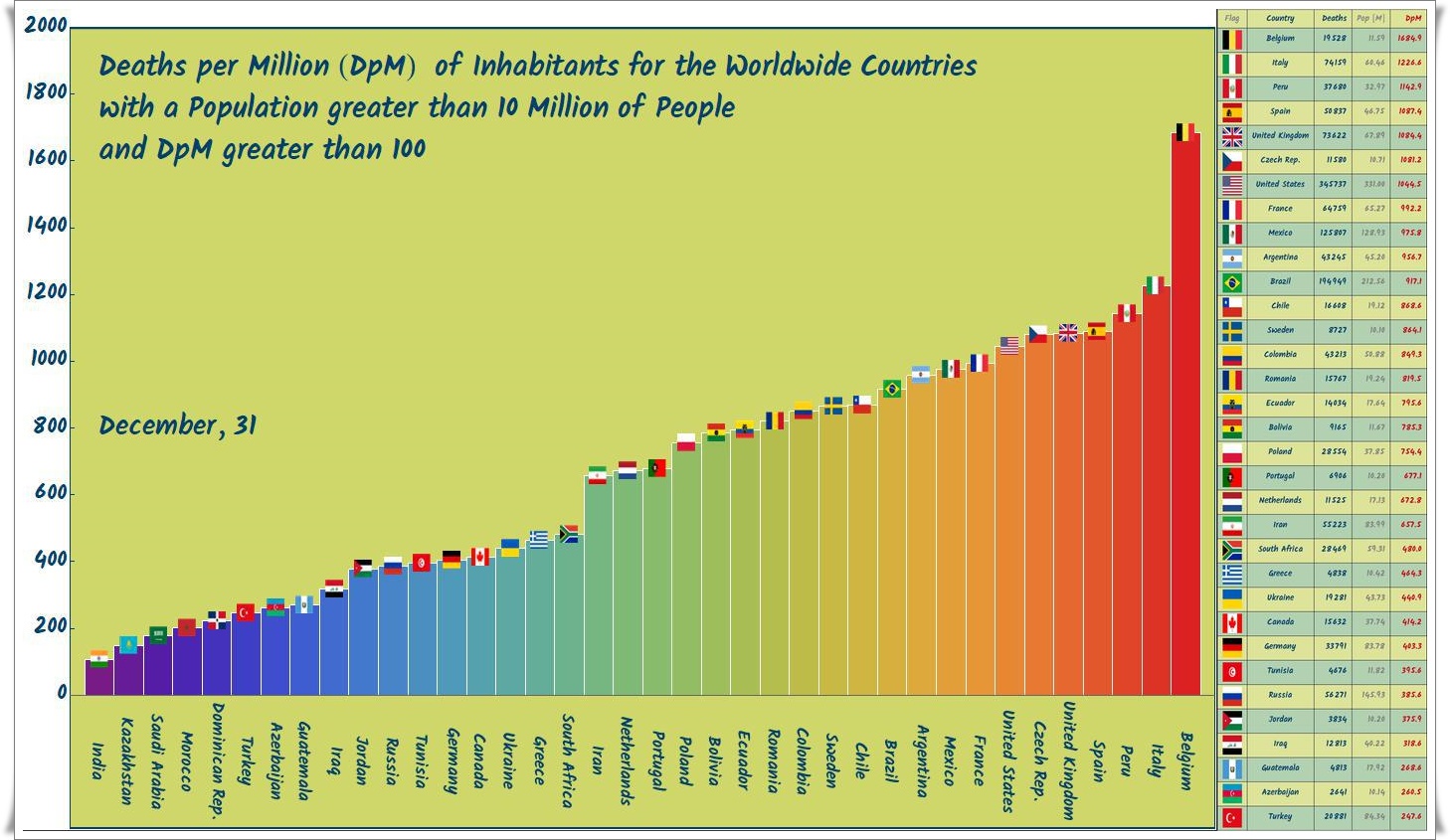}{\textbf{Worldwide deaths per million of inhabintants.}  The infograhic shows the deaths per million (DpM) of inhabitants of the countires with a population greater than 10 million of people  and DpM greater than 100 at December 31 (2020). In the attached Table, for the countries with DpM greater than 240,
we also show the absolute number of deaths.}}

\def\FigR{
\WideFigure{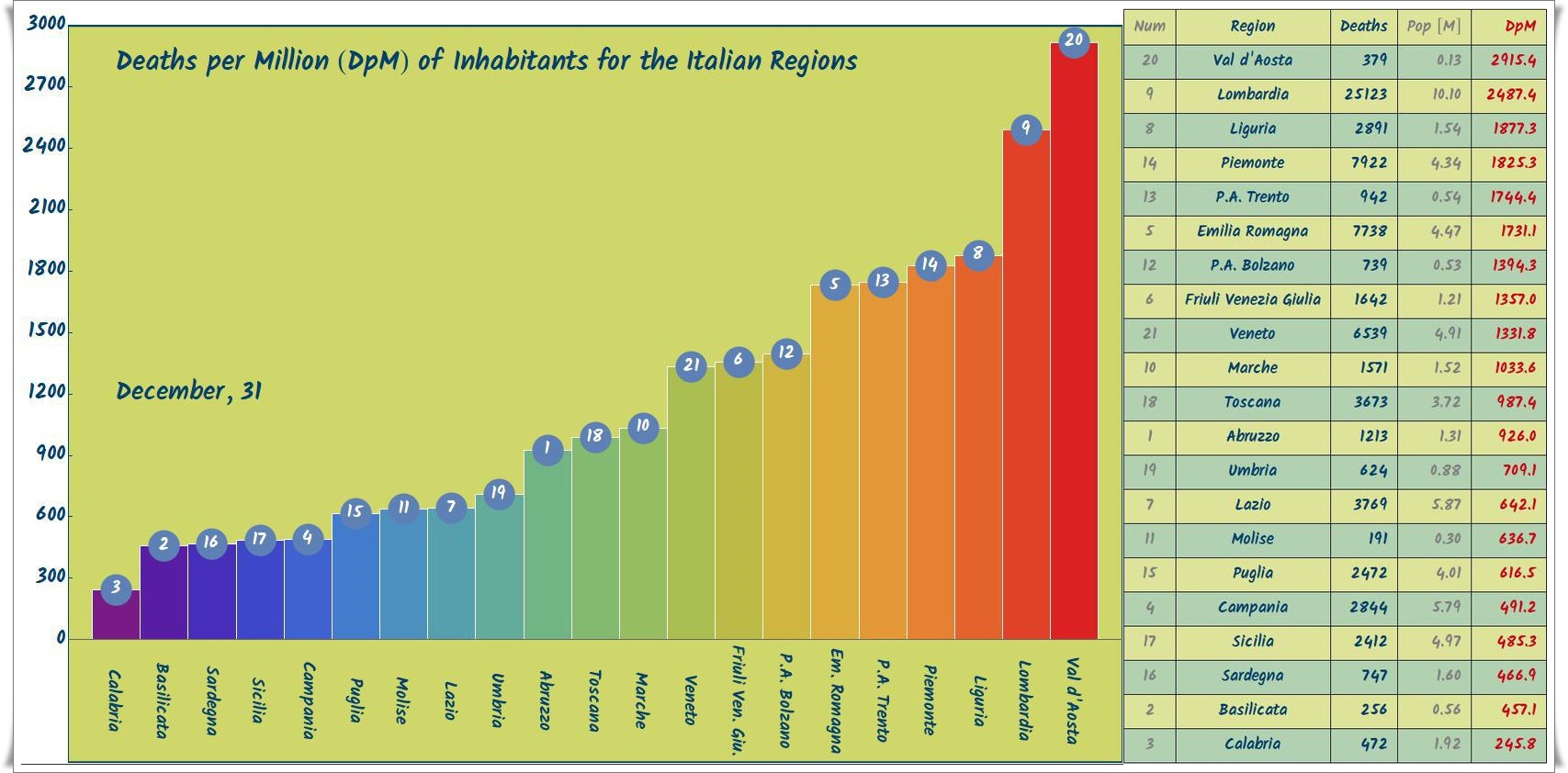}{\textbf{Regional deaths per million of inhabintants.} 
 The infograhic shows the deaths per million (DpM) of inhabitants of the regions and autonomous provinces of Italy
 at December 31 (2020). The absolute number of deaths appears in the attached Table.
}}

\def\FigA{
\WideFigure{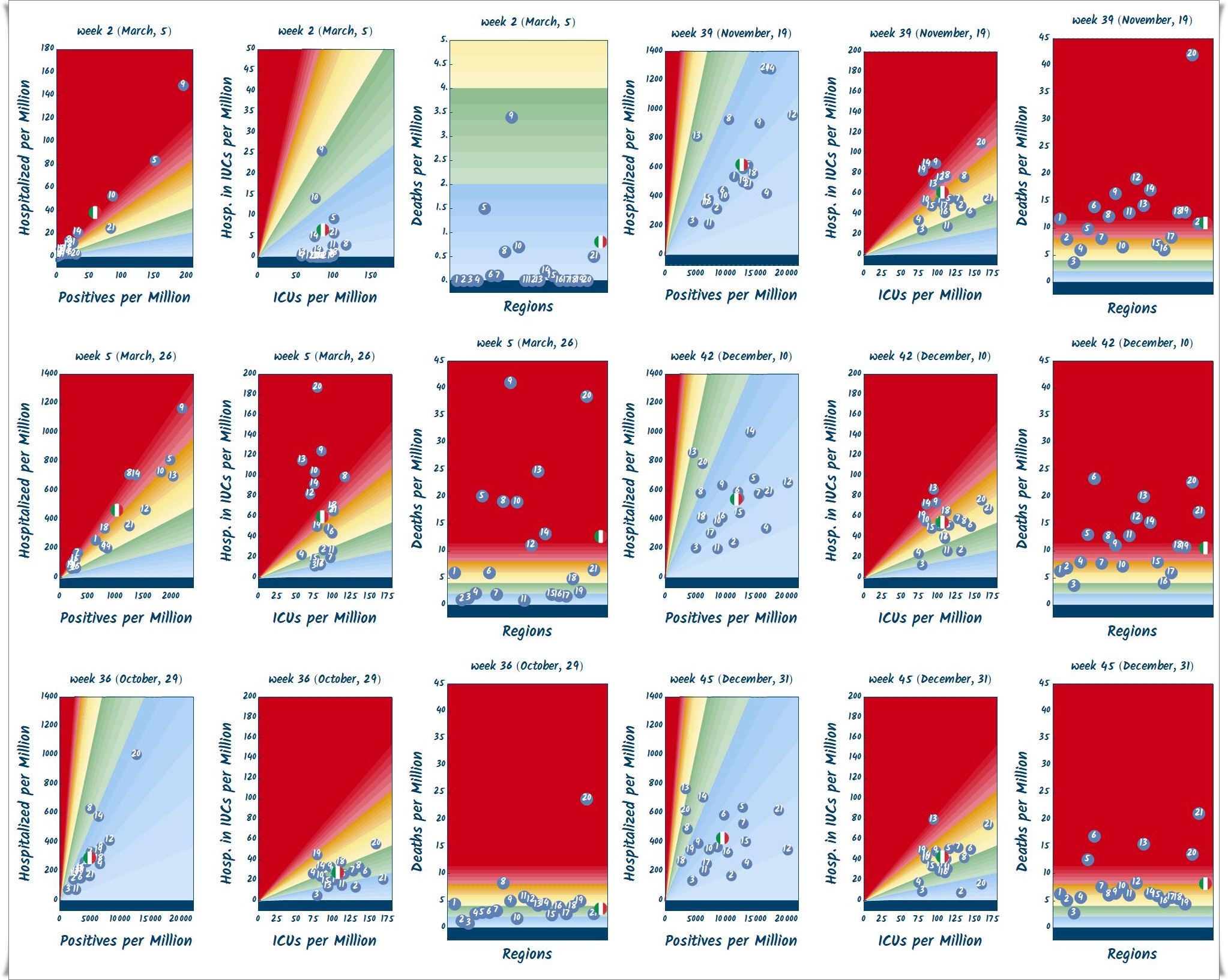}{\textbf{Pandemic plots.} The 7-day averages of hospitalized over positives, hospitalized in ICUs over ICUs, (B), and daily deaths are given for the pandemic week 2, 5, 36, 39, 42, and 45. 
}}

\def\FigB{
\WideFigure{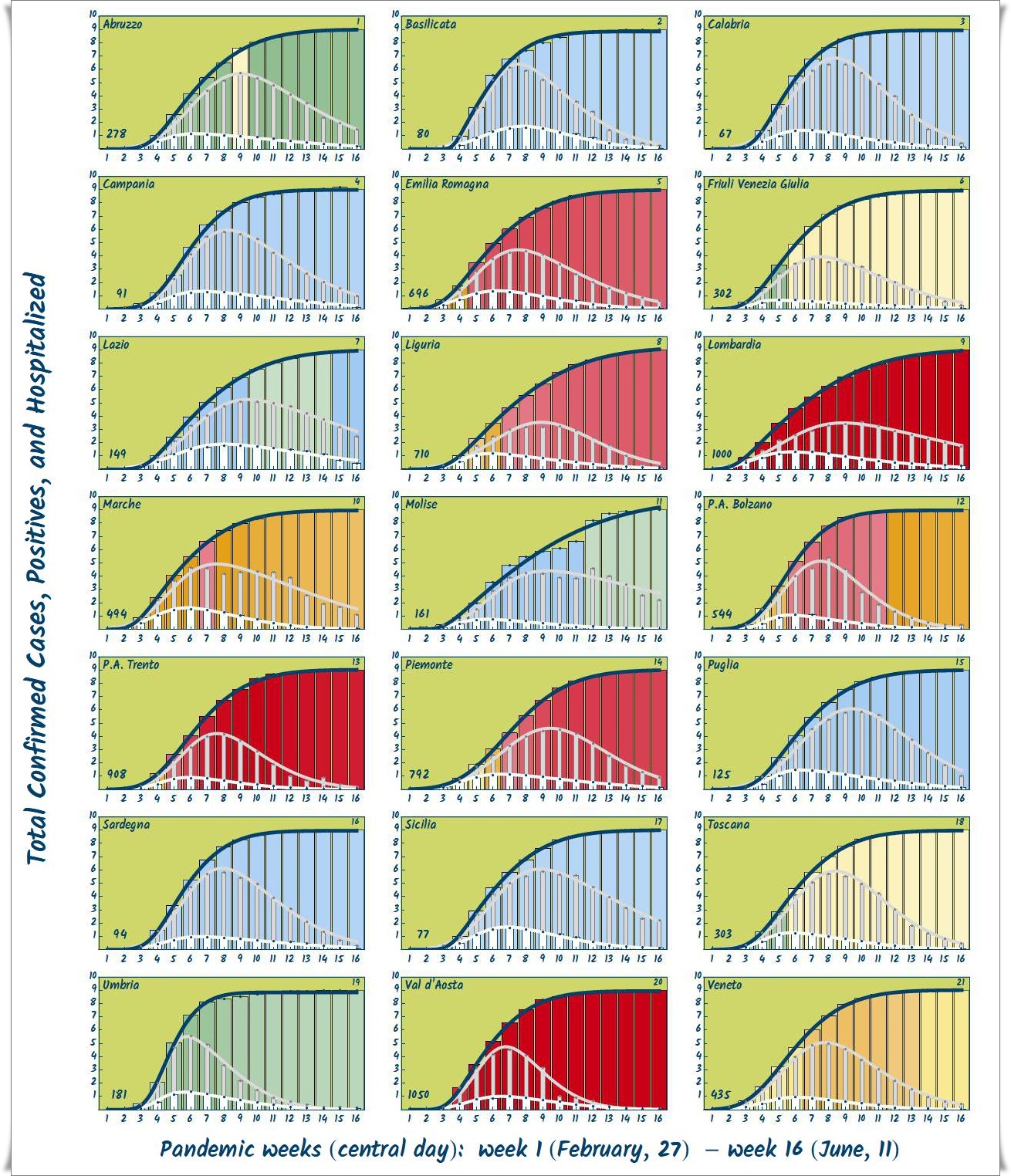}{\textbf{Skew-normal distributions.} The pandemic curves of total confirmed cases (colored histograms), positives (gray), and hospitalized (white) are modelled by skew-normal distribution. The analytical plots show an excellent agreement with the pandemic data.
}}

\def\FigC{
\WideFigure{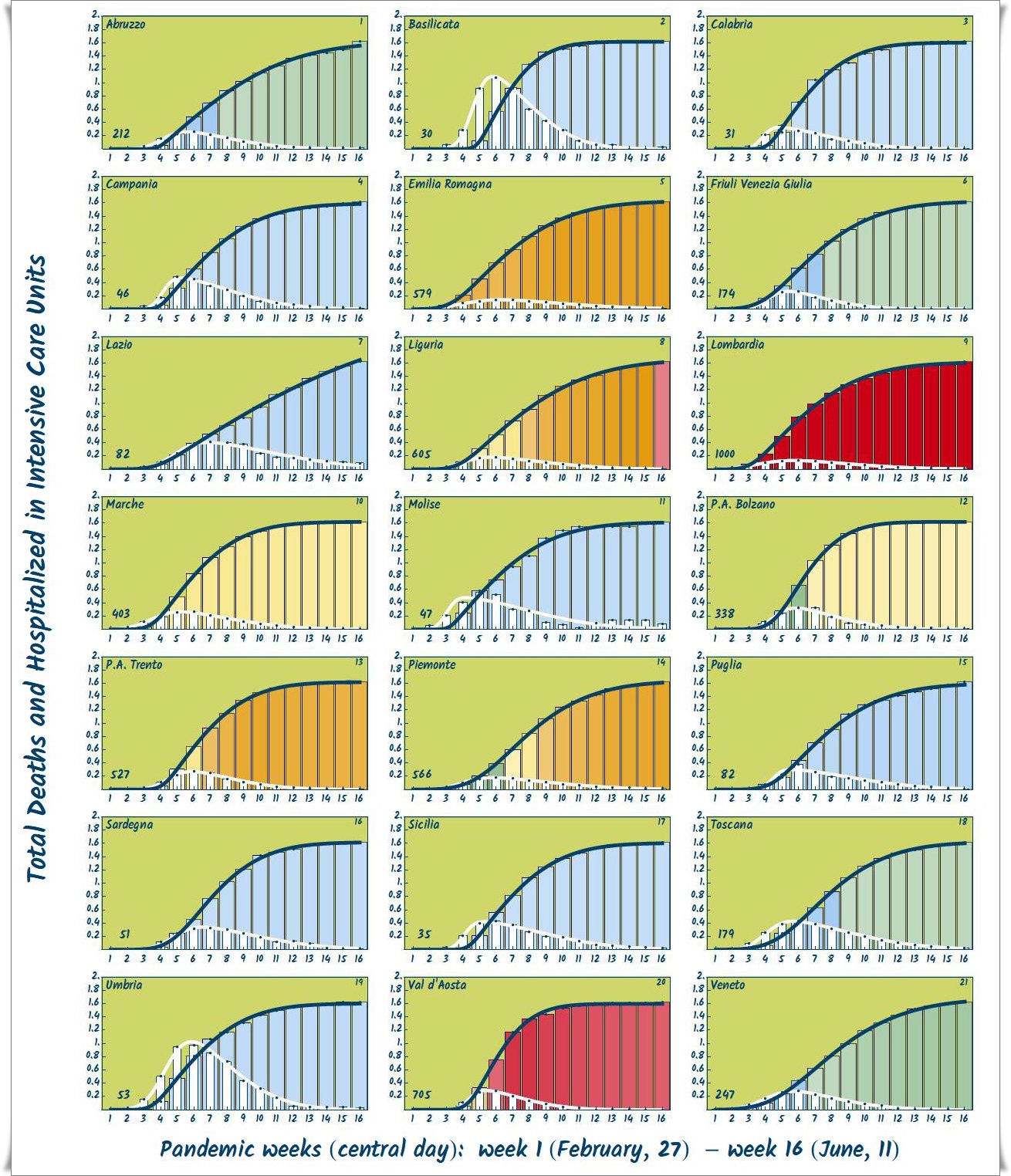}{\textbf{Skew-normal distributions.} The pandemic curves of total deaths (colored histograms) and  hospitalized in ICUs (white) are modelled by skew-normal distribution. The analytical plots show an excellent agreement with the pandemic data.
}}

\def\FigD{
\WideFigure{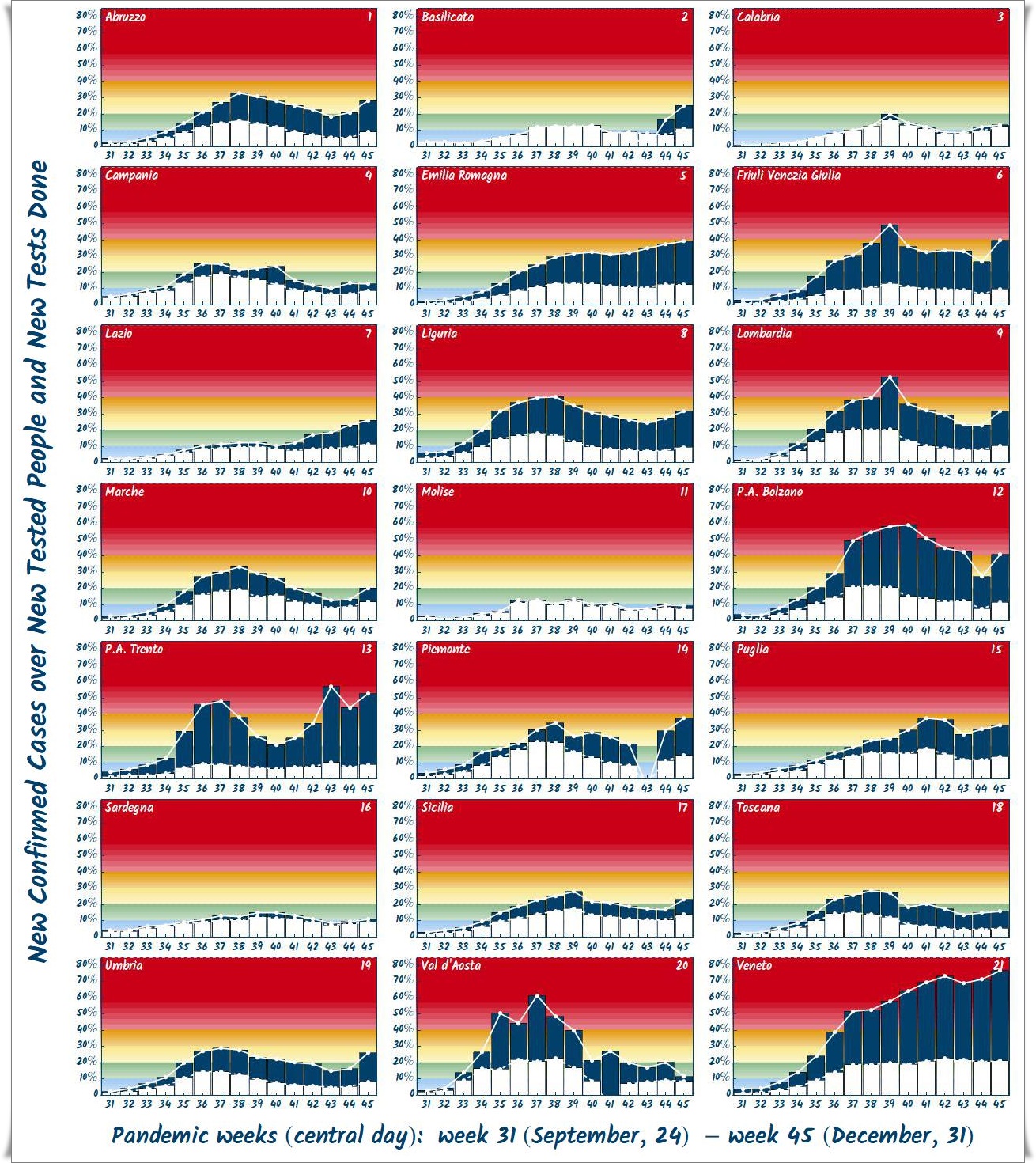}{\textbf{The first pandemic parameter.} The 7-day averages of new confirmed cases over new tests done (white histograms) and new people tested (blue histograms). The use of the incorrect ratio implies an underestimation of the infection reproduction number.   
}}

\def\FigE{
\WideFigure{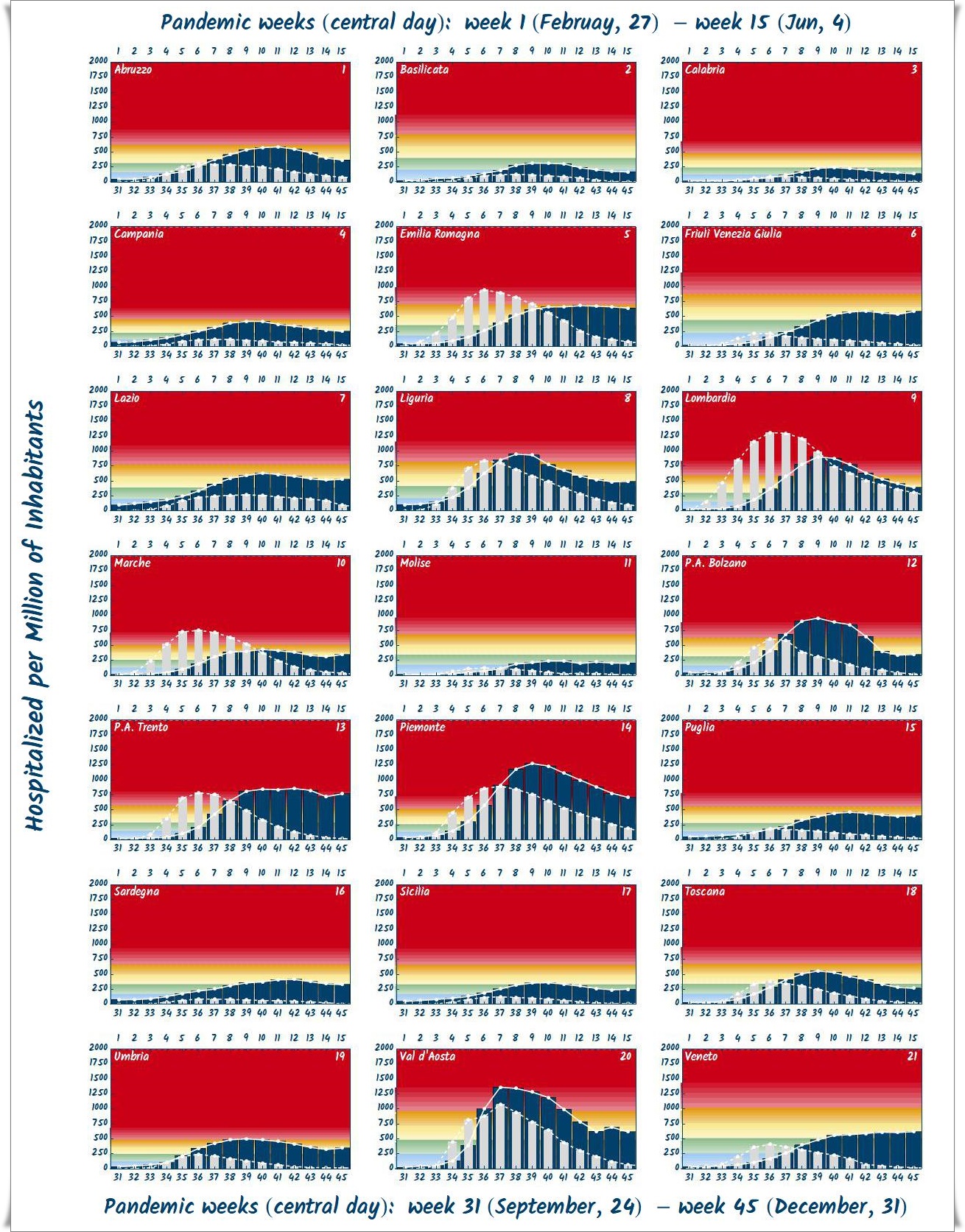}{\textbf{The second pandemic parameter.} The 7-day averages of hospitalized per million of inhabitants are plotted for the first 15 pandemic week (gray histograms) and for the last 15 ones (blue histograms).
}}

\def\FigF{
\WideFigure{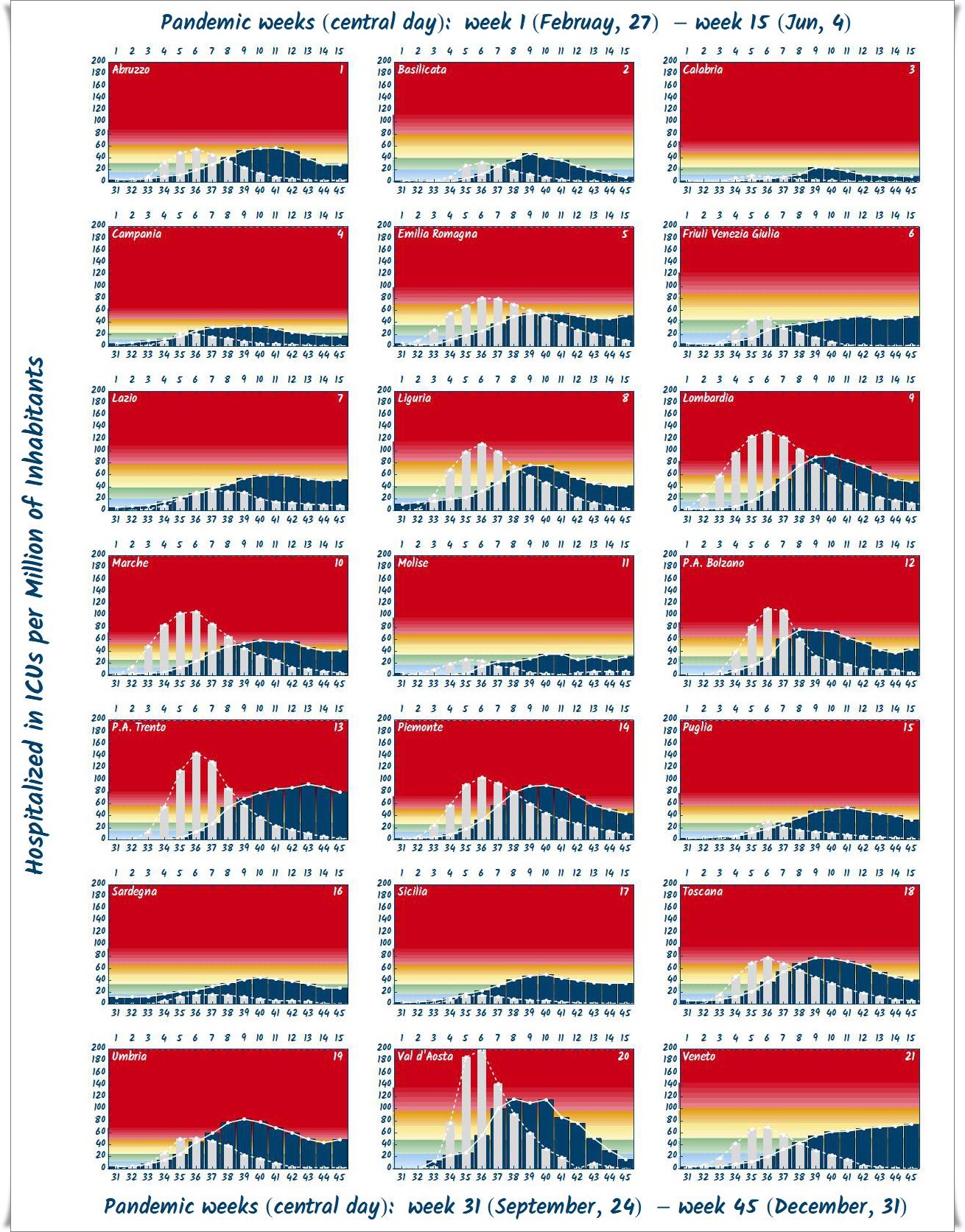}{\textbf{The third pandemic parameter.} The 7-day averages of hospitalized in ICUs per million of inhabitants are plotted for the first 15 pandemic week (gray histograms) and for the last 15 ones (blue histograms).
}}

\def\FigG{
\WideFigure{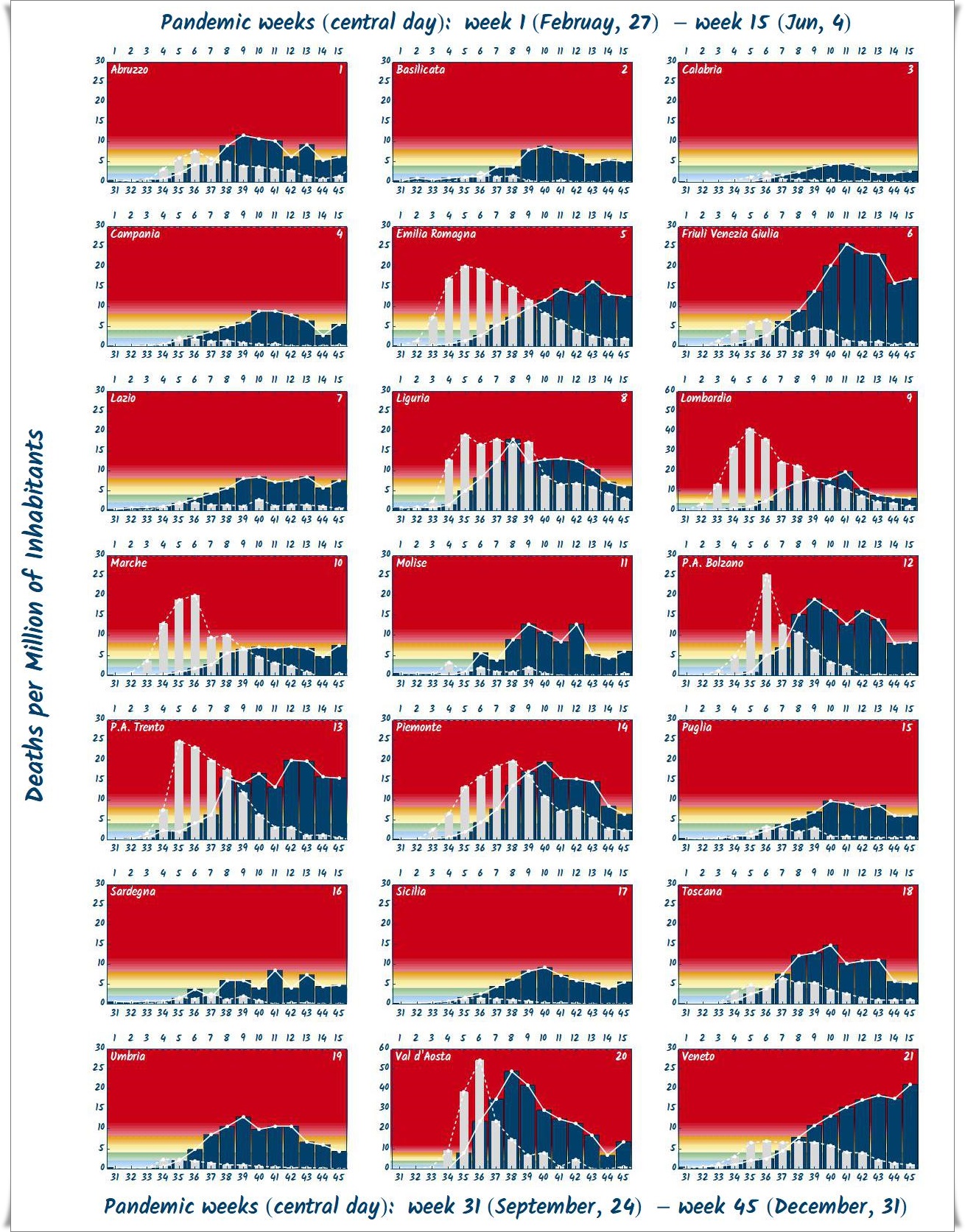}{\textbf{The fourth pandemic parameter.} The 7-day averages of daily deaths per million of inhabitants are plotted for the first 15 pandemic week (gray histograms) and for the last 15 ones (blue histograms).
}}

\def\FigH{
\WideFigure{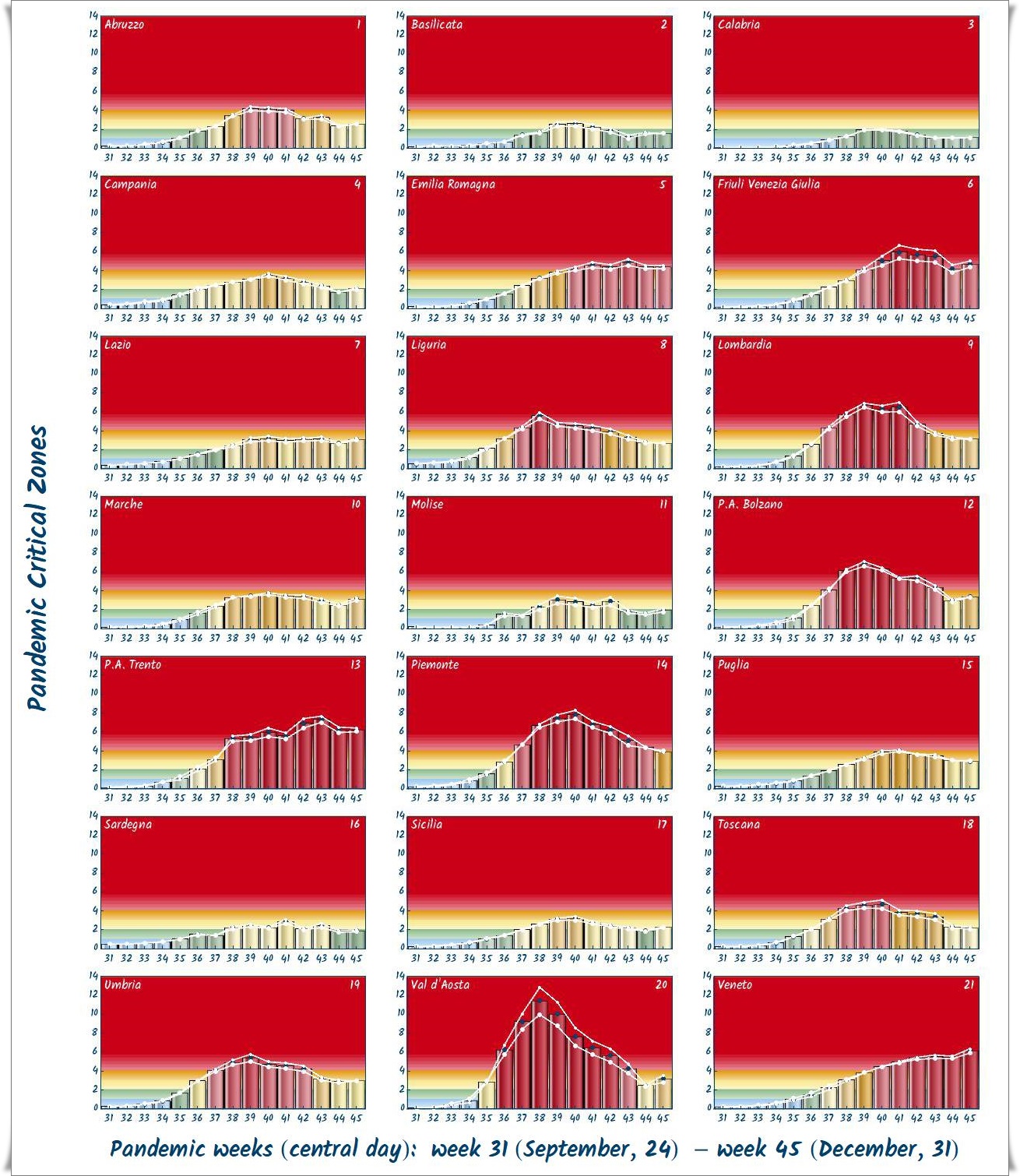}{\textbf{The pandemic criticality index (pci).} The pci time evolution during the last 15 week of the year is plotted for each Italian region.
}}

\def\TabA{
\WideTableSideCaption{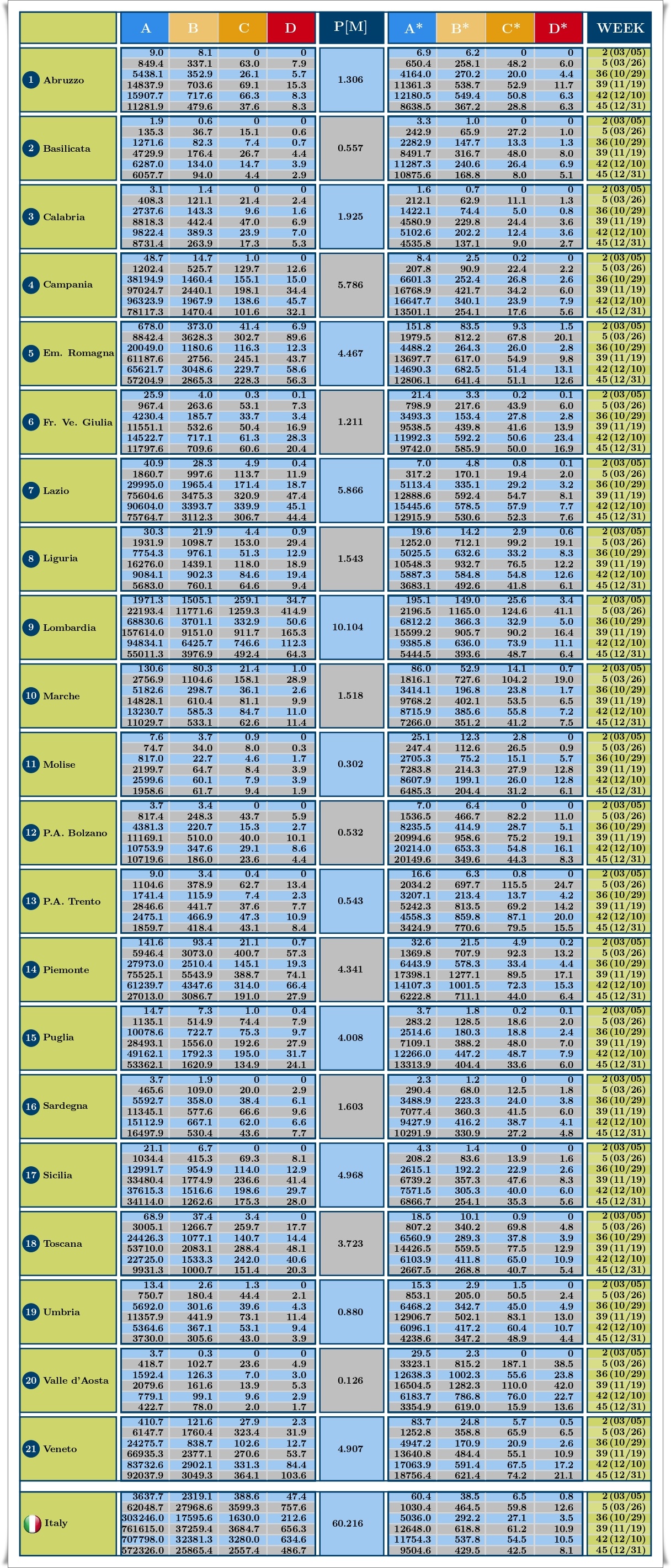}{\textbf{Pandemic data.} The 7-day averages of positives (A), hospitalized (B), hospitalized in ICUs (C), and daily deaths (D) are given for the pandemic week 2, 5, 36, 39, 42, and 45. In the Table, we also find the corresponding values per million  of inhabitants (A$^*$, B$^*$, C$^*$, and C$^*$). The regional populations appear in the central column, P[M].
}{24cm}}

\def\TabBC{
\WideTable{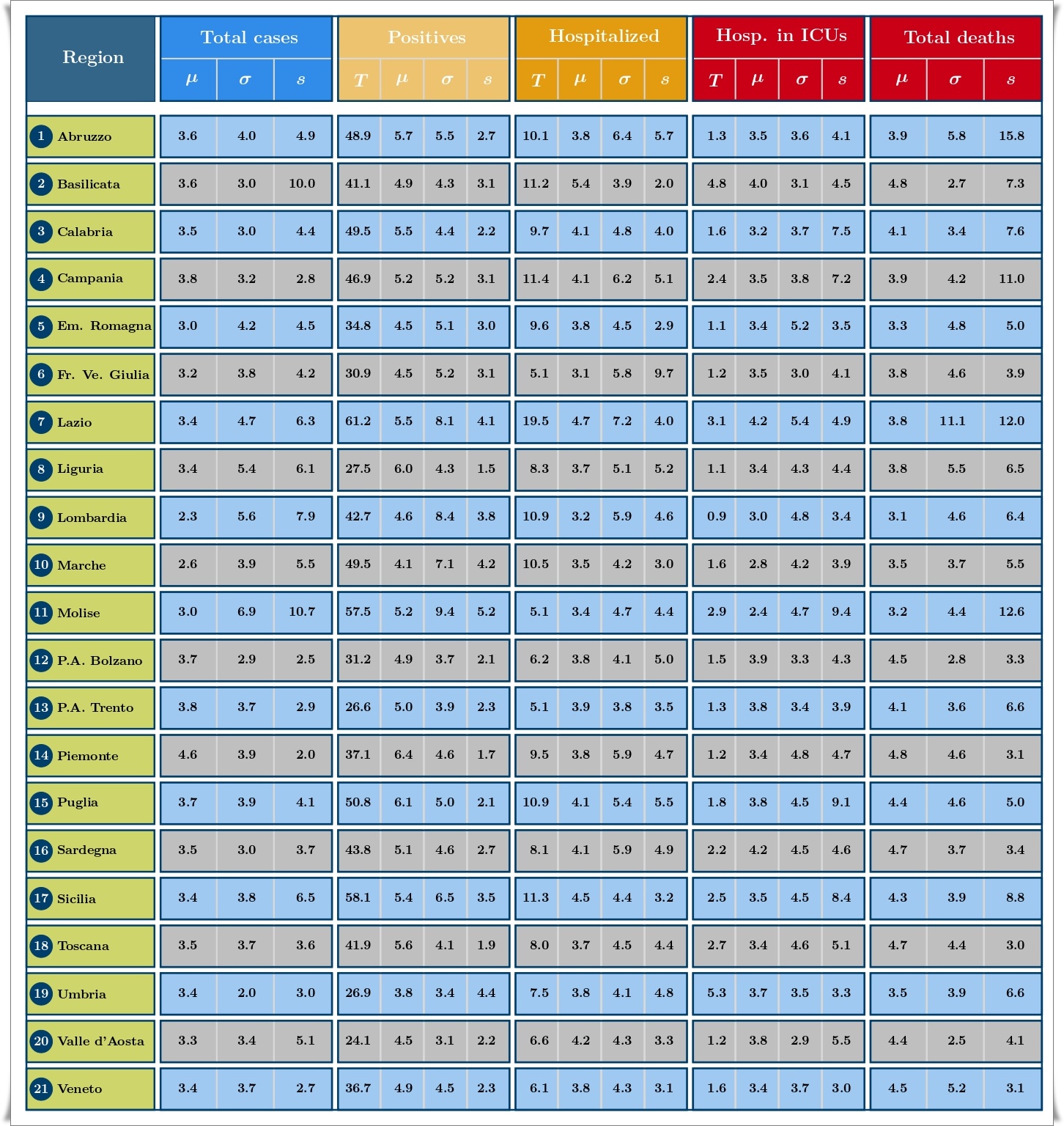}{\textbf{Skew-normal parameters.} The values of $T$, $\mu$, $\sigma$, and $s$ which allow to model the Italian pandemic curves are given for the total confirmed cases, positives, hospitalized, hospitalized in ICUs, and deaths. 
}}

\def\TabCol{
\WideTable{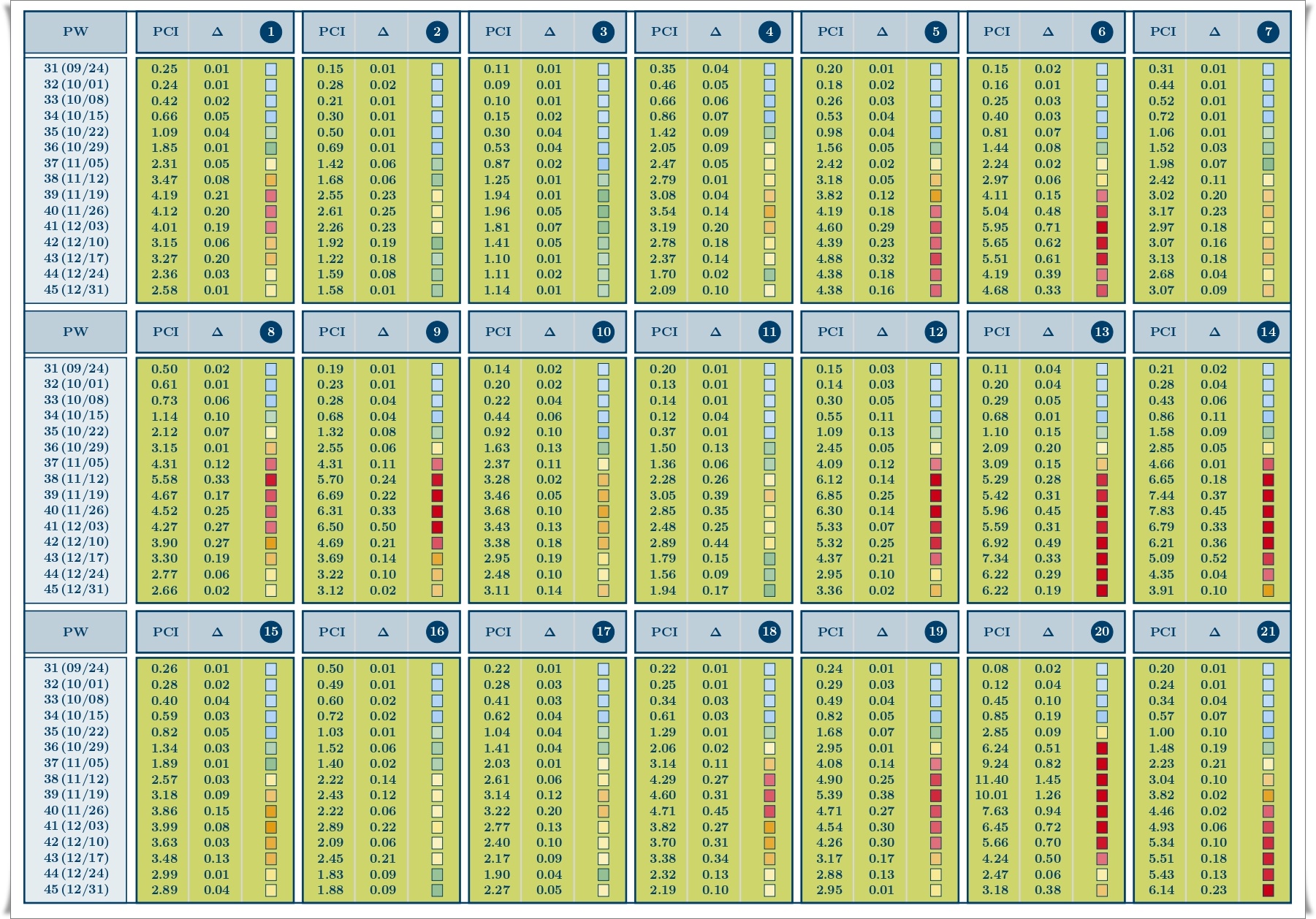}{\textbf{Numerical values of the pandemic criticality index (pci).} The numerical values of the regional pci  are given for the last 15 pandemic weeks of the year. In the Table, we also find the  color indicating the risk zone.
}}

\logo{
\colorbox{DarkGoldenrod}{\color{white}$\mathbf{\Sigma\hspace*{0.06cm} \delta \hspace*{0.04cm} \Lambda}$}
}

\journal{\textbf{\color{DarkRed} \normalsize
S. De Leo and M. P. Araújo [arxiv.org/abs/2102.03373 (q-bio.PE)]}}

\titlelines{5}
\title{A modelling study across the Italian regions:\\ 
Lockdown, testing strategy,  colored zones, and\\ 
skew-normal distributions. How  a numerical\\ 
index of pandemic criticality could   be\\ useful in tackling the CoViD-19.}
\newpage
\imgbgabstract{ 
As Europe is  facing the  second wave of the CoViD-19 pandemic, each country should carefully review how it  dealt with  the first wave of outbreak. Lessons from the first experience should be useful  to avoid  indiscriminate closures and, above all, to determine  universal (understandable)  parameters to guide the  introduction of containment measures to reduce the spreading of the virus. The use of  few (effective) parameters is indeed of extreme importance to create a link between authorities and population, allowing the latter to understand the reason for some restrictions and, consequently, to allow an active participation in the fight against the pandemic.
 Testing strategies, fitting skew parameters (as mean, mode, standard deviation, and skewness), mortality rates,  and  weekly CoViD-19 spreading data, as more people are getting infected,  were used to compare the first wave of the outbreak  in the Italian regions and to determine which parameters have to be checked before introducing  restrictive containment measures. We propose few \textit{universal} parameters that, once appropriately  weighed, could  be useful to correctly differentiate the pandemic situation in the national territory  and to rapidly assign the properly pandemic risk to each region.}

\author{
\names{Stefano De Leo and Manoel P. Araújo}
\affiliation{Department of Applied Mathematics, Campinas State University, Brazil}
\email{deleo@ime.unicamp.br}
}

\begin{document}

\sdlmaketitle

\section{Introduction}

Let's begin this article by analysing the data of deaths per million of inhabitants in the world as of December 31, 2020. In Figure 1, we find the data for the worldwide countries with a population greater than 10 million people and a number of deaths per million of inhabitants (DpM) over 100. In the attached Table, we also find the explicit data (absolute numbers of confirmed deaths, populations and  DpM) for the countries with a DpM greater than 240. The data can be looked up at the  
\textit{Coronavirus Source Data} by Our World in Data \cite{OWD}. Italy appears on the second place among the countries with the highest mortality rate with 1226.6 DpM, the number is obtained by dividing the absolute number of CoViD-19 confirmed deaths at December 31, i.e.  74159, by the Italian  population (60.46 million of inhabitants). The highest DpM number, 1684.9, belongs to Belgium. For this country, it is important to spend some words regarding this. In the beginning of the pandemic, for most countries  around the world, the CoViD-19 death toll was tallied from patients in the hospitals who tested positive for CoViD-19. Belgian authorities have gone further than that, by also including the deaths of non-hospitalized people who are suspected of having the virus, in particular the deaths in nursing (elderly)  homes. Since the beginning, Belgium is one of the few countries in Europe that have strictly followed and, maybe even extended the criteria for the CoViD-19 death classification, given by the World Health Organization:  ``A death due to CoViD-19 is defined for surveillance purposes as a death resulting from a clinically compatible illness, in a probable or confirmed CoViD-19 case, unless there is a clear alternative cause of death that cannot be related to CoViD disease (e.g. trauma). There should be no period of complete recovery from CoViD-19 between illness and death. A death due to CoViD-19 may not be attributed to another disease (e.g. cancer) and should be counted independently of pre-existing conditions that are suspected of triggering a severe course of CoViD-19''\cite{WHOd}. 
For example,  when Belgium reported 6262 deaths, 52\% of those fatalities were in nursing homes. Of these, only 4.5\% were confirmed to have had CoViD-19, with the rest just being suspected cases. It led Belgian CoViD-19 task-force spokesman and virologist \textit{Steven van Gucht} to suggest that when we  compare Belgium with other countries the Belgian death rate should be divided by two\cite{Reu}.  In Italy, the positivity to Sars-Cov-2 alone is not sufficient to consider death as due to CoViD-19, the presence of all the following four conditions is also necessary:
Death occurred in a patient definable as a microbiologically confirmed case (molecular swab) of CoViD-19; 
Presence of a clinical and instrumental picture suggesting CoViD-19; Absence of a clear cause of death other than CoViD-19;  Lack of full clinical recovery period between illness and death\cite{ISS}.   Looking at the 
ranking of countries by DpM  on June 28, Belgium, Spain, United Kingdom,  Italy and France appear, respectively, with 830, 606, 593, 574, and 456.  Considering the ratio of the reported CoViD-19 mortality and the excess mortality, 
the \textit{adjusted} deaths per million became 755 (Belgium), 1010 (Spain), 742 (United Kingdom), 857 (Italy), and 470 (France)\cite{EDM}. Clearly showing that,  during the first pandemic wave,  the CoViD-19 death toll was significantly  underestimated in Spain ($-$40\%), United Kingdom ($-$20\%), and  Italy ($-$ 33\%). 

In Figure 2, we  find an infographic  with the data for the Italian regions and the autonomous provinces. In the attached Table, we also find the absolute numbers of confirmed deaths, populations and  DpM. The data of the \textit{Protezione Civile Italiana}\cite{PC} are available for download at GitHub\cite{GH}. The regions of North-Western Italy (Val d'Aosta, Lombardia, Liguria, Piemonte) appear in the top 4, with a  36315 total deaths, which with a total population of 16.11 M implies  2254.2 DpM.  Immediately after these regions, we find the 3 regions (Emilia-Romagna, Friuli-Venezia Giulia, Veneto) and the 2 autonomous provinces (Bolzano, Trento) of North-East Italy  with 17600   total of deaths which with a total population of 11.66 M leads to 1509.4 DpM.   Going down the table, we find the 4 Central regions (Marche, Toscana, Umbria, Lazio) and one of Southern regions (Abruzzo) with 10850 deaths which with a total population of 13.3 M implies 815.8 DpM. Finally, the remaining Southern regions (Molise, Puglia, Campania, Basilicata, Calabria)  and the two islands (Sicilia, Sardegna) count  9405 total deaths, which with a total population of 19.15 M leads to  491.1 DpM. The geographical DpM difference amongst the Italian regions can be shown by using the red color for the regions with a $\mathrm{DpM}\geq 1800$, orange for $1200\leq \mathrm{DpM}< 1800$, yellow for  $600\leq \mathrm{DpM}< 1200$, green for $300\leq \mathrm{DpM}< 600$, and, finally, cyan for $\mathrm{DpM}< 300$.

The previous analysis introduces our next discussion. Why were the measures taken to tackle the first pandemic wave the same throughout the entire country? Did we have indicators that could avoid too restrictive measures for areas where the virus had not yet arrived? Did we have pandemic indicators to anticipate containment measures for the Northern regions where the virus had already spread for some time?
\begin{center}
\includegraphics[width=8cm, height=5cm]{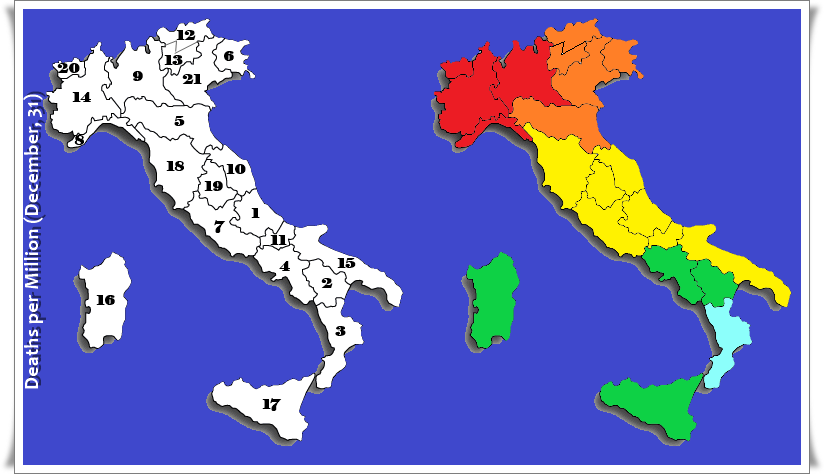}
\end{center}
The previous analysis introduces our next discussion. Why were the measures taken to tackle the first pandemic wave the same throughout the entire country? Did we have indicators that could avoid too restrictive measures for areas where the virus had not yet arrived? Did we have pandemic indicators to anticipate containment measures for the Northern regions where the virus had already spread for some time?

To answer the previous questions, let us see what containment actions the Italian authorities have adopted in the first 10 days of March 2020.  The first restrictive measures of prohibition of exit from and to access in some municipalities of Lombardy (Bertonico, Casalpusterlengo, Castelgerundo, Castiglione D'Adda, Codogno, Fombio, Maleo, San Fiorano, Somaglia, Terranova dei Passerini)  and a municipality of Veneto (Vo') were adopted by the Italian authorities with \textit{Decreto del Presidente del Consigli dei Ministri}  (DPCM) of the 1st of  March, DPCM200301 at \cite{DPCM}. Seven days later, the measure to avoid any movement of persons to and from these territories, as well as within  the same ones, except for movements  motivated by proven work needs, situations of necessity or health reasons were extended to the whole Lombardia and some provinces of Emilia-Romagna  (Modena, Parma,
Piacenza, Reggio Emilia, and Rimini), Marche (Pesaro and Urbino), Piemonte (Alessandria,
Asti, Novara, Verbano-Cusio-Ossola, Vercelli), and Veneto (Padova, Treviso, and  Venice), DPCM200308 at \cite{DPCM}. One day later, the measurements concerned the \textit{entire} national territory, DPCM200309 at\cite{DPCM}. A sequence of decrees in just a few days that clearly shows the panic of the  Italian authorities. Panic  probably due to the lack of  an adequate pandemic plan to deal with the outbreak. 

After January 11,  when the first CoViD-19 victim in China was confirmed, after January 13, the day on which  the virus caused the first death outside China (Thailand), after January 24, when the first cases in France were confirmed, after January 30, the date on which the WHO declares the CoViD-19 a \textit{global health emergency} \cite{Global}, Italian authorities  should have immediately activated the pandemic  plan by worrying, for example, about having a large supply of personal protective equipments, the  careful monitoring of lung diseases compared to previous years and the control of intensive care units in the country. 

In the excellent report \cite{Zambon}  prepared by the CoViD-19 Emergency Team at the WHO European Office for Investment for Health and Development in Venice (Italy), the authors  discussed  why the first phase of Italy's response to the CoViD-19 brought its health system to near collapse creating a panic in the population. In February, life in Italy had not changed much: carnival festivities, tourists in the cities, people in  the ski resorts and  football fans at the stadium. A scary example  was  the UEFA Champions League match played at the \textit{Giuseppe Meazza} stadium of Milano between the Italian team of Bergamo, \textit{Atalanta}, and the Spanish one of \textit{Valencia} with 45792 spectators. 
According to an analysis conducted by INTWIG\cite{INT}, in collaboration with Report\cite{Rep} and BergamoNews\cite{BNs}, among the 3400 (of 36000) Atalanta fans interviewed more than 20\% had symptoms compatible with CoViD-19 in the 15 days following the match. This football match was a clear pandemic bomb that then made Bergamo become a symbol of the epidemic that devastated Lombardia  between February and March 2020. A study of the Institute \textit{Mario Negri} (Bergamo) and of the Deparment of Biomedical and Clinical Science, University of  Milano \cite{EBioM}  estimated the cumulative prevalence of SARS-CoV-2 infection in Bergamo in a group of workers who returned to the workplace after the end of the Italian lockdown on 5th May 2020. Performing an enzyme-linked immunosorbent assay (ELISA)   to detect the humoral response against the spike and nucleocapsid proteins of  SARS-CoV-2, as well as  nasopharyngeal swabs to assess the presence of SARS-CoV-2 using real-time reverse transcription polymerase chain reaction (rRT-PCR), the researchers  observed the prevalence of SARS-CoV-2 infection in the province of Bergamo reached 38.5\%, significantly higher than has been reported for most other regions worldwide. By using the result of this study, we can also calculate the \textit{real}  Infection Fatality Rate for the outbreak in Bergamo.    The IFR is one of the important numbers alongside the herd immunity threshold, and has implications for the scale of an epidemic and how seriously we should take a new disease. Assuming that the 38.5\% of the Bergamo population was infected by  the virus, the IFR in this area reduces from 20\% (comfirmed deaths over confirmed  infected people) to almost 1\% in perfect agreement with  the interval between 0.5\% and 1\% calculated across different countries\cite{IFR}. Observe that for a common flu, the IFR is around 0.1\%. Just a factor 3 (i.e. 0.3\% instead of 0.1\%) without a vaccine is sufficient to lead to a collapse of the health systems of most countries in the world.
 
  The study we aim to present in this report is based on the analysis of 45 pandemic weeks. The central day of the weeks, around which we will calculate our 7-day averages is Thursday, with the first week centred on February 27 and the last one on December 31.  Before presenting the analysis of pandemic data, it is convenient to spend a few words on the situation of the intensive care units (ICUs) in Italy before the first and second pandemic waves. Intensive care units, providing treatment for people who are in a very critical situation,  are staffed with specially trained healthcare professionals, and contain sophisticated monitoring equipment. Due to the fact that some people infected by CoViD-19  may be unable to breathe on their own, they need to use a machine   that helps with breathing (a ventilator) and  monitoring equipment  to measure important body functions, such as heart rate, blood pressure and the level of oxygen in the blood. The collapse of the intensive care units surely was and still is  the main preoccupation in facing this pandemic. We observe that the collapse of the health system is not only due to the lack of intensive care beds but also to the lack of trained healthcare professionals (this was one of the main problems in Italy). 
The total number of beds in ICUs at the beginning of the outbreak was  5179. In Summer,  the commissioner (Domenico Arcuri) responsible for the implementation and coordination of the measures necessary for the containment and contrast of the epidemiological emergency CoViD-19  prepared a plan to create 3553 additional ICUs. On October 13, the call for companies closed, and the start of the implementation work 
was scheduled  for the  end of October.  Too late, because at that time the second wave in Italy had already arrived. In the meantime, the regions added a total of 1279 stable beds to their initial 5179 and so the current dowry  is 6458 places \cite{24a}. This implementation was done  with significant regional differences also due to the accumulated delays: The regional plans were expected at the end of June and instead were only approved at the end of July.   The situation of intensive care beds for the Italian regions at the beginning of the first and second pandemic waves is shown in the following graphics
 \begin{center}
\includegraphics[width=8cm, height=5cm]{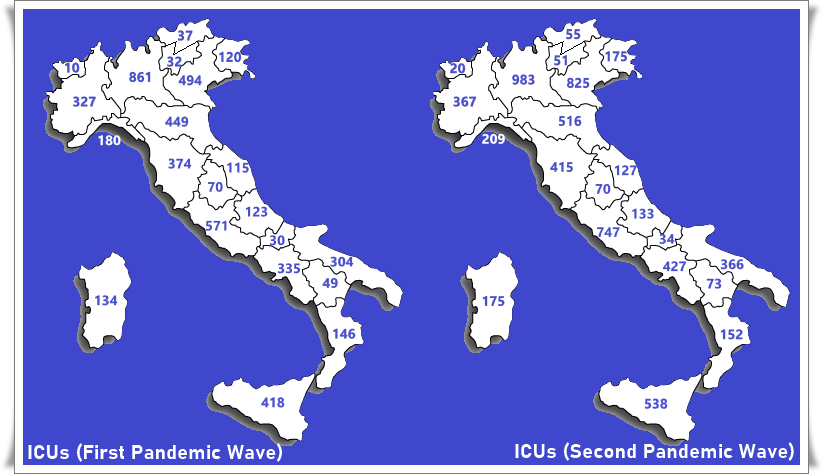}
\end{center}
The number of ICUs suggested by the World Health Organization is  150 per million of inhabitants. To understand the  critical situation in Italy at the beginning of the first pandemic wave,  we will highlight with the red color the regions with a number of intensive care beds less than 75, with orange  those with a number of beds greater than or equal to 75 but less than 100, with yellow greater than or equal to 100 but less than 125, with green greater than or equal to 125 but less than 150, and, finally, with cyan greater than or equal to 150.  
 \begin{center}
\includegraphics[width=8cm, height=5cm]{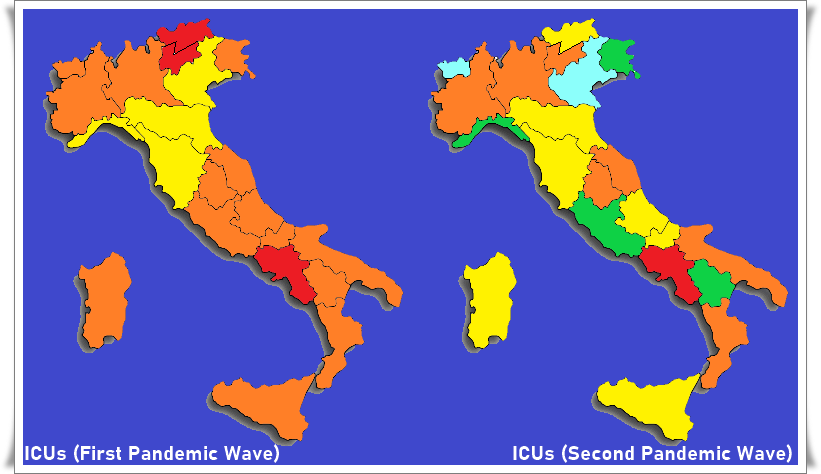}
\end{center}
Even after the implementation of new intensive care beds, only 4 regions, Friuli Venezia Giulia (144.5 ICUs per million), Basilicata (131.1), Liguria (135.5), and  Lazio (127.3),  appear in green and 2, Veneto (168.1) and Val d'Aosta (158.7) in cyan. Campania presents the most critical situation with 73.8 ICUs per million of inhabitants. Notwithstanding this critical situation (red zone), at December 31, Campania with a DpM of 491.2 (green zone) appears in the lower part of the regional DpM Table and at the beginning of the infographic shown in Figure 2.

\section{Delayed action and lockdown}

As observed in the Introduction, the Italian government, in an attempt to stop the spreading of the first wave of the pandemic,  adopted strict containment measures (lockdown) throughout the whole national territory starting from March 9. For the second wave, on the other hand, the Italian authorities  applied the regional division into colored zones of pandemic risk.

In Table 1, we find, region by region, the number of the  daily averages for  positive (column A), hospitalized (B), hospitalized in ICUs (C) and  deaths (D).  In this Table, we also find the population (P) of each region and the numbers of daily averages for  positive (A$^*$), hospitalized  (B$^*$), hospitalized in ICUs  (C$^*$), and deaths  (D$^*$)   per million of inhabitants. The data of Table 1 were used to prepare Figure 3.  To facilitate the understanding of data and tables, we will analyse some specific cases. Let us consider the situation of Lombardia in the second pandemic week (centred at March 5). In this week, the daily average of positives, hospitalized, hospitalized in intensive care, and deaths was  1971.3, 1505.1, 259.1, and 34.7. To compare Lombardia with the other regions, we normalise these number by its population (10.104 M) obtaining: 195.1, 149.0, 25.6, 3.4. One of the most important pandemic parameter (often underestimated during the first pandemic wave) is the ratio of  hospitalized over positive people. Indeed, when a minimum threshold of hospitalized  (20 per million) is reached, the hospitalized/positives  ratio  is the first indicative number of the gravity of the disease.   The more advanced the stage of infection is the more diffusion we have, so  it becomes more likely that weaker groups will be affected  and  more people will need hospitalization. The more people are hospitalized the more people could complicate their clinical status with the possibility of needing intensive care. In the second pandemic week, Lombardia had 149.0 hospitalized per million of inhabitants and 195.1 positives to CoViD-19 leading to a scary  76.4\%. In Figure 3, we have chosen the cyan color when this ratio is less than 10\%, green between 10\% and 20\%, yellow between 20\% and 30\%, orange between 30\% and 40\%, and, finally, red when the ratio is greater than 40\%. Other indicators such as  hospitalized in ICUs/number of ICUs and deaths per million clearly confirm the Italian government's responsibility  in the lack of  firmness and determination in curbing the spreading  of the pandemic in the Northern regions, in particular Lombardia. The CoViD-19 occupancy of  ICUs  is another important indicator. In Figure 3, we have chosen the  cyan color when this ratio is less than 15\%, green between 15\% and 30\%, yellow between 30\% and 45\%, orange between 45\% and 60\%, and, finally, red when the ratio is greater than 60\%.  For the daily deaths per million, the colored zones were fixed at 2, 4, 6, and 8. At this stage of the outbreak, due to  the lack of pandemic planning  and territorial monitoring 
the CoViD-19 deaths are obviously   underestimated. All the regions appear in the cyan zone (the only one in green is Lombardia).  

On March 8, the Italian authorities  imposed  restrictions  to Lombardia and some provinces of Emilia Romagna, Marche, Piemonte, and Veneto on March 8 and, 24 hours later, the restrictions were extended to the entire national territory on March 9. An  incomprehensible way of handling the first wave of the pandemic and leading to panic in the Italian population.  From the data of the second pandemic week (Table 1) graphically shown  in Figure 3, it is clear that the containment measures for  regions such as  Lombardia (9), Emilia Romanga (5), Marche (10),   Piemonte (14), and Veneto (21) should have been  taken  much earlier. This is confirmed 3  weeks later (fifth pandemic week centred at March 16), when the situation precipitated not only for the mentioned  regions but for all the other  Northern regions. For some regions the  7-day averages of deaths per million reached scary values: 9-Lombardia (41.1), 20-Val d'Aosta (38.5), 13-P.A. Trento (24.7), 5-Emilia Romagna (20.1), 8-Liguria (19.1), 10-Marche (19.0), 14-Piemonte (13.2), and 12-P.A. Bolzano (11.0). It is important to observe that a  number of 7-day averages deaths per million near to 10 means that we have at least 100 CoViD-19 hospitalized in ICUs per million of inhabitants and this puts in checkmate  most of the health systems in the world. The main success in tackling the pandemic is to flatten all the pandemic curves  by adopting timely containment measures, by increasing the number of intensive care units, by engaging new  medical staff, by choosing appropriate drugs to minimize hospitalization, and when possible by using vaccination. The daily deaths per million for the Northern regions of Italy is a clear evidence of  the untimely action by the authorities and the lack of protocols that could indicate the right path for people who, remaining confined at home, could have health problems not only related to CoViD-19. It is clear that it is easy to judge   the action of the Italian government after observing and studying the pandemic results, but, with all possible justifications for the unforeseeable situation, it is also clear that the lack of a pandemic plan and the delayed action of the authorities played a fundamental role in the disaster seen in the Northern regions, in particular in Lombardia. 

The word \textit{quarantine} has its origin  in Venice in the 14th century during the Black Death. The city decided to extend the quarantine period for ships and people from 30 to  40 days. This was a very  important (fortuitous) decision.  The incubation period for the bubonic plague was estimated to be 37 days \cite{SD}.   The national quarantine was the \textit{only} answer found by the Italian government.  The toll that a 10-week block during the first pandemic wave took on the Italian people was a very costly one, both from an economic and social point of view. The Central, Southern regions and Islands were forced to a strict lockdown notwithstanding  the virus had  practically arrived not yet entered their territory (we shall come back to this point later when we model the pandemic curves of the first wave by skew-normal distributions). The panic created in the population  led people with CoViD-19 to aggravate their health before seeking medical treatment and also created disastrous effects in the prevention of other diseases.

The deaths in the first half of 2020 could at least have been honoured if the second wave of the pandemic was addressed properly. But even when facing the second wave, the Italian authorities have shown a poor ability not only in communicating with the population but, above all, in fighting the pandemic.

\section{The 21 pandemic parameters}

In the absence of a vaccine or effective drugs treatment, and because of the still low level of immunity in the population, a rapid resumption of sustained transmission may occur in the
community (second wave) once we reduce the containment measures, adopted to tackle  the first pandemic wave. Trying to avoid this, the Minister of Health with a decree of April 30\cite{MS} indicates the monitoring criteria that must be followed in order to promptly classify the pandemic risk   and to be able to adopt local containment measures
to face  a new outbreak wave. The monitoring included the following indicators:\\

6 indicators of data collection capacity (IA);

6 indicators of diagnostic control, investigation and contact tracing (IB);

9 indicators of  transmission stability and  resilience of the health system (IC).\\

\noindent The 21 indicators are explicitly given in Appendix. 
An excessive number of indicators, typical of  the Italian bureaucracy, is clearly, independent of their validity or not, an additional difficulty in monitoring the disease. The regions can probably (but always with great difficulty) monitor these indicators 
when the outbreak gives a respite but clearly 21 indicators cannot be monitored and even more understood by the population during the worst phase of the pandemic.   To understand the great confusion generated in the population by the Italian authorities population,  we recall  some of the incomprehensible decisions taken by the authorities through dozens and dozens of decrees each of one with a number of pages between  100 and 150.  During the CoViD-19  crisis,  it is necessary to communicate with the population in a simple, clear, and effective way exactly the opposite of what the Italian authorities did. After the first Summer months with open discotheques and without the need to use personal protective equipment outdoors (in many European countries the use of mask outdoors  was imposed in the beginning of the reopening period) the Minister of Health on August 16 decides to impose  the use of mask outdoors. But, only from 6pm to 6am of the day later and only when agglomerations could not be avoided\cite{MS}. Certainly a confused and unclear way of communicating with the population who clearly do not understand why at 6pm the use of the mask is triggered and 5 minutes before it is not. A good example of how authorities shouldn't communicate with their own population. Only two months later, the use of the  mask outdoors  is compulsory during the whole day, DPMC201013\cite{DPCM}.   The use of the mask also has a psychological effect, keeping the attention of the user higher.

Before analysing the measures taken by the Italian government to face  the second pandemic wave, let us observe the current situation in the pandemic week number 36 (centred on October 29). In this week the Italian 7-day average of daily positives and hospitalized was 761615.0 and 37259.4, respectively, leading to an hospitalized/positives ratio of approximatively 5\% and a (7-day average) daily deaths per million of 3.5 (the absolute number was 212.6), see Table 1 and Figure 3. The daily hospitalized/positives ratio (5\%) confirms that we are  already  in the second pandemic wave and  the daily deaths near to 4 that we  should have already taken containment measures. In the final sections, we shall discuss in detail the numerical  \textit{pandemic criticality index} which permits to correctly classify the risk of each region. For the moment, let us have a look at the  (7-day average)  daily deaths per million of inhabitants. At week 36: 1-Abruzzo (4.4), 8-Liguria (8,3) , 9-Lombardia (5.0), 11-Molise (5.7), 12-PA Bolzano (5.1), 
13-PA Trento (4.2), 14-Piemonte (4.4), 16-Sardegna (3.8), 18-Toscana (3.9), 19-Umbria (4.9) and 20-Val d'Aosta (23.8) overcome the national mean, with two regions (Liguria and Val d'Aosta in a very critical situation).

At November 4, The Minister of Health gave the first colored classification of the Italian regions\cite{MS}. Four regions appear in the red zone (Lombardia, Piemonte, Val d'Aosta, and Calabria), two in the orange one (Puglia e Sicilia), and the remaining regions and  Bolzano and Trento in the yellow zone.
 \begin{center}
\includegraphics[width=5cm, height=6cm]{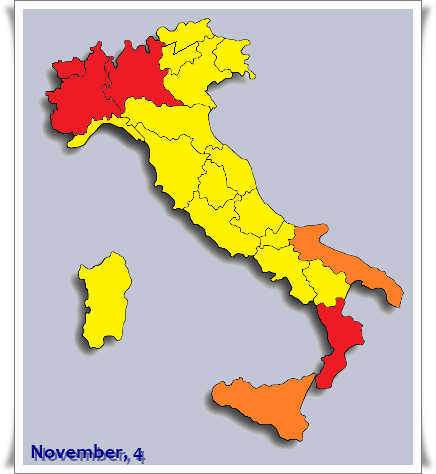}
\end{center}
Why the yellow color was assigned to Liguria and to Calabria the red one? In the week before the classification (week 36), Liguria  had a  ratio of hospitalized/positive inhabitants of 976.1/7754.3$\,\approx\,$12.6\%, a CoViD-19  ICUs occupancy rate of 
51.3/209$\,\approx\,$24.5\%, and a (7-day average) daily deaths per million  of 
12.9/1.543$\,\approx\,$8.4 (the difference from 8.3 which appears in Table 1 and used in our previous discussion is due to the approximation done in 12.9).  In the same week, the  hospitalized/positive ratio, the ICUs rate, and the daily deaths per million  for Calabria were  143.1/2737.6$\,\approx\,$5.2\%,  9.6/152$\,\approx\,$6.3\%, and 1.6/1.925$\,\approx\,$0.8.     

From the data of Table 1 and their graphical presentation in Figure 3, the best numbers belong to Calabria. This region was the first one to be put in the red zone and never it was collocated in the yellow one.  For Veneto, it was always determined the yellow zone. See the following picture.
 \begin{center}
\includegraphics[width=8cm, height=14cm]{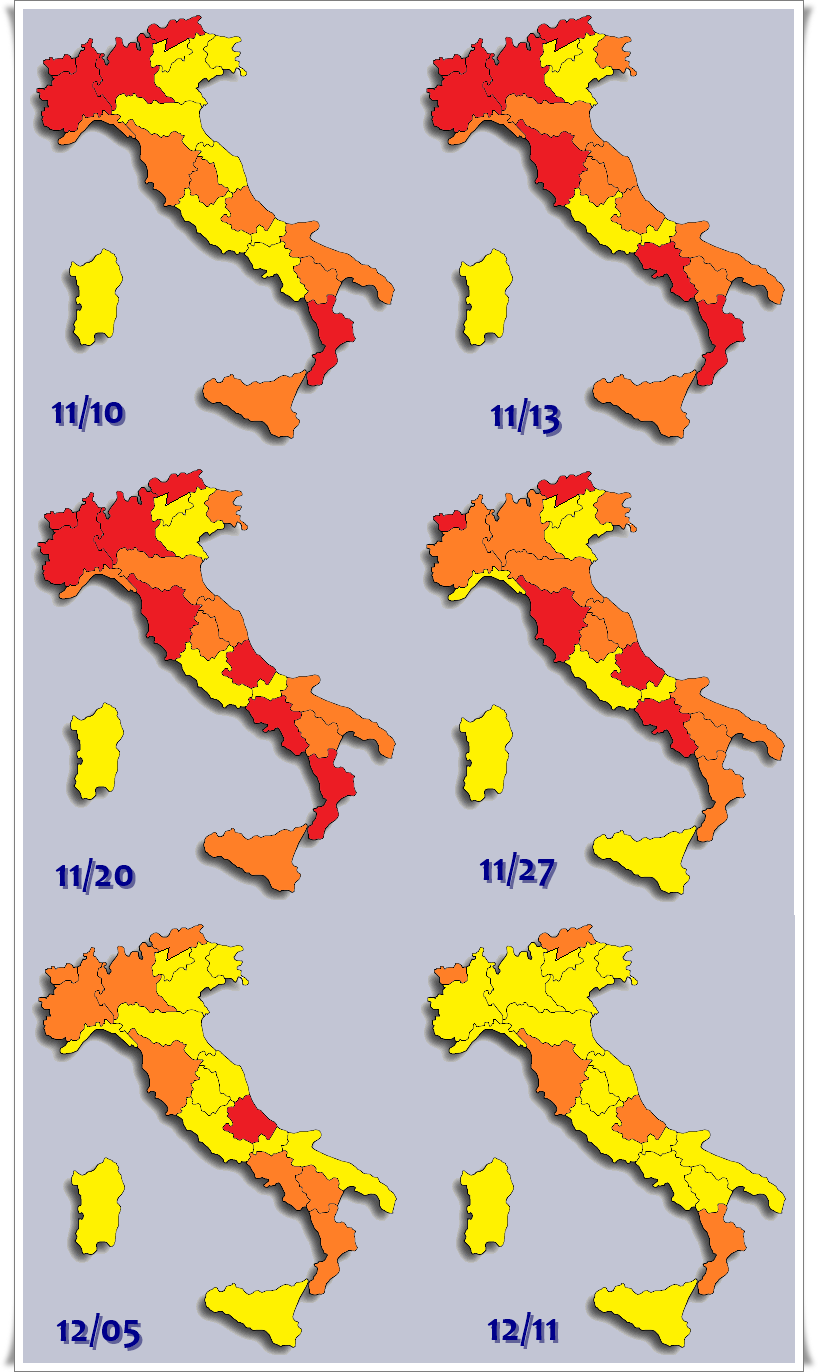}
\end{center}
Let us examine, its number in week  39 (centred at November 19) and 42 (December 10). The CoViD-19 ICUs rate  passed from 32.8\% to 40.2\% and the daily deaths from 9.1 to 17.2. Nothing was done to really face the spreading of the disease  in this region (remember that the yellow zone is the lighter pandemic risk zone).  Consequently, in the last week of the year, we find Veneto in a very scary situation with a  CoViD-19 ICUs rate
of 44.1\%  and a (7-day average) daily deaths per million of 21.1.

The great confusion  in determining  the pandemic gravity by the Italian authorities was the initial motivation of our work.  In this paper we aim to propose a numerical index of pandemic criticality weighted by a few effective parameters: The ratio of new infected people over new tested cases, the number of  hospitalized per million of inhabitants, the  CoViD-19 ICUs rate, and, finally, the daily deaths per million.

\section{Skew normal distributions}

In studying the CoViD-19 data, we must use two distinct approaches\cite{MED01,MED02}. First, until reaching  the curve peak, we have to model the data by a Gaussian distribution \cite{ND}. Once the peak has been reached, we have to introduce a parameter which models the  \textit{asymmetry} of the curve. In this case the  skew-Gaussian distributions\cite{SND,SNDa,SNDb,SNDc} come into play.  The explicit analytical formula of the skew probabilities density functions, used in this paper  to fit  the Italian pandemic curves of the pandemic first wave, is given by
\begin{eqnarray}
\mathcal{PDF}_{\hspace*{-0.1cm}a}(x)&=&\frac{T_a\,
\exp\left[ -\,\displaystyle{\frac{(x-\mu_a)^{^{2}}}{2\,\sigma_{a}^{^{2}}}}  \right]}{\sqrt{2\,\pi}\,\sigma_{a}}\,\,\times \nonumber  \\ 
 & & 
\mathrm{Erfc}\left[ -\,\frac{s_a\,(x-\mu_a)}{\sqrt{2}\,\sigma_a}   \right]
\end{eqnarray}
where  
\begin{eqnarray*}
a& = & \textrm{confirmed cases per Million of Inhabitants (pMoI)}, \\
& = & \textrm{positives pMoI}, \\
& = & \textrm{hospitalized pMoI}, \\
& = & \textrm{hospitalized in ICUs  pMoI}, \\
& = & \textrm{deaths pMoI}, 
\end{eqnarray*}
and $\mathrm{Erfc}$ is the complementary error function,
\[
\mathrm{Erfc}(z)\,\,=\,\,\frac{2}{\sqrt{\pi}}\,\int_{_{z}}^{^{\infty}}\hspace*{-0.25cm}\mathrm{d}t\,\,\exp[\,-\,t^{^{2}}\,]\,\,.
\]
The skewness of the distribution,  defined by
\begin{equation}  
\gamma\,\,=\,\,\left(2\,-\,\frac{\pi}{2}\right)\,
\left(\,\frac{\delta}{\sqrt{1\,-\,\delta^{^{2}}}}\,\right)^{^{3}}
\end{equation}
where
\[\delta\,\,=\,\, \sqrt{\frac{2}{\pi}}\,
\frac{s}{\sqrt{1\,+\,s^{^{2}}}}\,\,,\]
has a value in the interval  $\left(\,-\,1\,,\,1\,\right)$.  The mean value of the distribution is obtained  by
\begin{equation}
\textrm{mean}\,\,=\,\, \mu \,+\,\sigma\,\delta\,\,.
\end{equation}
The mode (maximum) has not an analytic expression but, as shown in \cite{SNDd},  an accurate closed form is given by
\begin{eqnarray}
\textrm{mode}&=& \mu\,+\,\sigma\left(\,\delta\,-
\,\frac{\gamma\,\sqrt{1\,-\,\delta^{^{2}}}}{2}\right.\nonumber \\
& & \left. \,\,\,\,\,\,-\,\,\mathrm{sign}(s)\,
\exp[\,-\,\frac{2\,\pi}{s}\,]\,\right)\,\,.
\end{eqnarray}
For total confirmed cases and deaths, we used
the cumulative skew-normal distributions,
\[
\mathcal{CDF}_{\hspace*{-0.1cm}a}(x)\,\,=\,\, 
\displaystyle{\int_{_{-\,\infty}}^{^{\,\,x}}\hspace*{-0.2cm}\mathrm{d}
\widetilde{x}\,\,\,\,\,\mathcal{PDF}_{\hspace*{-0.1cm}a}(\widetilde{x})}\,\,.\]
The four parameters, $T$, $\mu$, $\sigma$, and $s$, were calculated by using  
the 7-day averages data and by fitting them by the \textbf{\texttt{NonlinearModelFit}} of  the computational program Wolfram Mathematica \cite{WMat}. The fitting parameters for total confirmed cases and deaths, for daily positives, hospitalized, and hospitalized in ICUs are found in Table 2.

In order to optimize the  graphical presentation of the pandemic data, we have decided to normalise  the total confirmed cases and  deaths to those of Lombardia. At the sixteenth pandemic week (centred on June 6), Lombardia (number 9 in the plots) reached 
90979.7  (7-day average) total confirmed cases   and 16374.9  (7-day average) total deaths. Numbers that, considering the population of 10.104 M, lead to 9004.3  
confirmed cases per million of inhabitants and 1620.6 deaths per million. For Lombardia, 
we used the scale factor 1000.  For total confirmed cases and deaths, all the other regions have been then normalised to the Lombardia values.  The regional scale factor appears in the lower left corner of each regional plot,  see Figure 4 and  5. Let us explain how the scale factors have to be used. In the pandemic week 16, Calabria (num.\,3), Puglia (num.\,15), and  Veneto (num.\,21) had  1160.4, 4513.4, and  1920.1 (7-day averages) confirmed cases and 97.0, 529.4, 1966.1  (7-day average) deaths. By considering their populations, 1.925 M, 4.008 M, and 4.907 M, we find  602.8,   1126.1, and 3912.8 total confirmed cases per million of inhabitants and  50.4, 128.6, and  400.7  deaths per million.  By using the scale factor appearing in the corresponding plots (Figure 4 and 5), we can obtain the  total confirmed cases and   total deaths for each region. In particular, for the regions cited above, we find  
\[
\begin{array}{lclcrcr}
\textrm{Calabria} &:& 9.0 & \times& 67 &=&603\,\,,\\    
                  & & 1.6    & \times& 31 &=&50\,\,,\\  
 \textrm{Puglia} &:& 9.0 & \times& 544 &=&1125\,\,,\\    
                  & & 1.6    & \times& 82 &=&131\,\,,\\     
\textrm{Veneto} &:& 9.0 & \times& 435 &=&3915\,\,,\\    
                  & & 1.6    & \times& 247 &=&395\,\,.                                            
\end{array}
\]
The scale factors can also be  used to determine the regional colors for the total confirmed cases and deaths of the first pandemic wave: cyan at 150, green at 300, yellow at 450, orange at 600, red at 1000, with a proportional color gradation between two colors. Looking at the total confirmed cases per million of inhabitants: 7 regions appear in the cyan zone, 5 of which (Basilicata, Calabria, Campania, Sardegna, and Sicilia)  with a scale factor of deaths per million less than 100 and two of which (Lazio and Puglia) with a number between 100 and 150, and 2 regions (Molise and Umbria) in the green zone. For the total deaths per million of inhabitants, we have 9 regions in the cyan zone, all of them (Basilicata, Calabria, Campania, Lazio, Molise, Puglia, Sardegna, Sicilia, and Sardegna) with a scale factor less than 100, and 4 regions in the green zone (Abruzzo, Friuli Venezia Giulia, Toscana, and Veneto). From Figure 4 and 5, it is clear that the critical numbers belong to the Northern regions with the exception  of Friuli Venezia Giulia and Veneto. 

The skew-Gaussian distributions (blue, gray, and white lines)  show an excellent agreement with the pandemic data (colored, gray, and  white histograms). The fitting parameters can then be used to compare different regions and, for the same region, to compare  positives, hospitalized, and hospitalized in the intensive care units data.  It is also important to note that the gradation of colors in the pandemic  curves of confirmed cases and  deaths gives us  an idea of the temporal evolution  in each region. Normalisation is always done with respect to Lombardia.

In Lombardia, for the first pandemic wave,  the mortality rate seems to be 1.6/9=17.8\%. It is obvious that this rate cannot be the \textit{real} infection fatality rate  (IFR) of Sars-CoV2 because 17.8\% is calculated by using the number of known confirmed cases and  not the number of \textit{real} infected people. In an interesting study done in a small German town,  of 919 individuals  15.53\% were infected.  By applying  this infection rate to the total population in the community, i.e. 12597,  we can estimate  1956  (real) infected people. Having 7 Sars-CoV2 deaths reported by the local authorities, the \textit{real} IFR can be estimated to be  7/1956 =0.36\% \cite{NatC}.  By using the  infection rate of 15\% in Lombardia, we should have, during the first wave, 1.5 million of infected people. This means a factor 16.7 with respect to the number of confirmed cases reported  by the local authorities. In this case, the \textit{real} IFR for Lombardia is found around 1\%. Lower than    17.8\% but still greater than 0.36\%. Observe that a factor 2.5 in the total  deaths implies 16000 instead of 6400 
deaths.  This high IFR, very close to the one found for Bergamo\cite{EBioM}, is  clearly  due to the confusion generated in the population and to the lack of a national pandemic plan. In a panic situation, people to do not seek medical help at the right time and, in many cases, this  aggravates their health conditions generating a difficult situation to be managed by the medical staff. We also remember that during  the first wave, the Lombardia elderly homes  were even used to accommodate patients with CoViD-19. An incomprehensible choice that brought the the virus to the most fragile age range with obvious dramatic consequences. In the Northern regions the containment measures were adopted too late, but this was not the only cause of
the disaster.  The enormous difficulties in which health professionals worked clearly show, as observed in the excellent report by Zambon et al. \cite{Zambon},  the lack of a  pandemic plan.  The World Health Organization was accused of conspiring with the Italian authorities to remove the Zambon's report revealing the mismanagement of Italian government  at the beginning of the CoViD-19 pandemic. In the 102-page report, the authors observed that the  national  pandemic plan had not been updated since 2006 and that, due to being unprepared, the initial response from hospitals was \textit{improvised}, \textit{chaotic}, and \textit{creative}. 
For the Southern regions, where the virus had not yet arrived or had only reached a minimum extent and therefore could be controlled with targeted measures, the lockdown  was identical to that adopted for the Northern regions. As we will see later, things have not changed much when the second pandemic wave arrived. A shame for the Italian government and a sadness for its population who certainly did not deserve such treatment and who, contrary to what the national authorities did, responded with discipline and determination in the months of March, April, and May. The pandemic curves were controlled by the discipline and determination of the Italian population and by the heroic work of health professionals which compensated for the confusion and  the  incompetence and the lack of programming of the Italian authorities which found in a medieval lockdown the \textit{only} answer to the first wave of pandemic.

\section{The four pandemic parameters}

In this Section, we turn our attention on 4 pandemic parameters: New confirmed cases over new tested people, hospitalized and hospitalized in ICUs per million of inhabitants, and ,finally, daily deaths. The first parameter is the one used to determine the infection reproduction number, but this cannot be the only parameter to be observed, as the other 3 parameters can be more effective in understanding the real situation of the pandemic.

CoViD-19 tests can be useful to reduce the virus diffusion by using timely  preventive isolation measures and by monitoring close contacts of infected people. Before discussing the importance of a massive testing strategy, let us see which types of tests are currently being used\cite{CoT}. The most effective test to detect the presence of the Sars-CoV-2 virus is the one based on the molecular analysis. This test has been used in Italy to identify people who have contracted the virus. Once a person carrying the virus has been identified, it is clear that a first measure is the isolation of the same and, once isolated, the local authorities have to check  his close contacts.  Digital  proximity tracking tools are used to widen the network of possible contacts.  The effectiveness of such digital tools clearly depends on a high coverage and utilization rate among the population. National lockdown during the first pandemic wave, lack of a pandemic plan, confusion, bad communication, and  questionable choices in the reopening period (such as the non-obligation to wear a mask outdoors) led the population to underestimate the possibility of a second pandemic wave. The contact tracing \textit{app} suggested by the Italian government was, due to the very low acceptance within the population, practically useless. 
The lack of an effective network of  regional contact tracing led  to the  collapse when the pandemic began to spread within the territory. Without a massive testing strategy and without an effective contact tracing is clear that many asymptomatic people  will never be  identified.
In addition to the molecular test, other tests should have been employed: The quick swab antigen test, the classic and  rapid serological test, and the salivary test. For the quick swab antigen test, the sample collection methods are similar to those of molecular tests (nasopharyngeal swab). This test has a lower sensitivity but it allows to identify the antigens of the SARS-CoV-2 virus in  a very short time (about 15 minutes).  The serological (or immunological) test detects the presence of specific antibodies that the immune system produces in response to infection (IgA, IgM and IgG) and their quantity in the blood. They tell us if we have contracted the virus and for how long. This test requires a venous blood sample, and is carried out in specialized laboratories. The rapid serological test is based on the same principle as the classic one, but it only tells us whether or not specific antibodies for the virus are present in the body. The average response time is about 15 minutes and can also be done outside the laboratories. Serological tests, by their  nature, are unable to tell whether the patient has an ongoing infection, but only whether or not he has come into contact with the virus. They can therefore provide useful information to understand how many people have come into contact with the virus (stratify them by age and geographical region) and to determine if a natural (herd)  immunity has been achieved. Salivary collection is simpler and less invasive than nasopharyngeal swab or blood sampling. This type of  test assesses the presence of the virus in the body and it could be  very useful for screening large numbers of people. So, when performed on a recurring basis (every 72 hours), it could allow for rapid isolation and outbreak control decisions. This test is, for example, very important in reopening schools.

The choice to do not prioritize a massive use of tests for screening was certainly one of the problems regarding the territorial pandemic control, unfortunately, it was not the only mistake. In the beginning of a pandemic, having a few number of infected, the differences between new tests done and new people tested   is so small to be practically  insignificant.  Once the pandemic is spreading enough in the territory, the rate of  new cases over new tests done (see the regional white histograms in Figure 6) can lead to wrong conclusions. What is to be  considered, it is  the number of new cases over the number of new people tested (blue histograms in Figure 6).

Let us consider an explicit example to understand the great confusion created by Italian authorities when looking at the pandemic data and in the communication with the population. 
First of all, let us  show the importance of using weekly averages instead of daily data. At November 29, Puglia had the following daily numbers: 907 (53218-52311) new confirmed cases and  8285 (780364-772079) new tests done. From these data,  the daily infection rate communicated to the population was  10.9\%. On day later, Puglia had 1102 (54320-53218) new cases with 4151 (7845515-780364) new tests done and, consequently, an infection rate of 26.5\%. This frightening leap get into panic midia and population. It is clear that a 7-day average, it is a more appropriate way to treat the pandemic data. By using, the 7-day averages, the rate passes from 15.9\%, (53218-43507)/(780364-719303), to 16.0\%,  (54320-44487)/(784515-61434), see the Puglia white histogram at the week 40 in Figure 6.  After understanding that in the communication with the population is better used 7-day average, let us now analyse which was the mistake done by the Italian authorities in facing   the second pandemic wave.  As observed before,  to calculate the \textit{real} infection rate, we have to use the number of new tested people and not the number of new tests. By using the number of new tested people, we find  30.2\%, (53218-43507)/(538195-506049), and  29.3\%,  (54320-44487)/(541174-507599), see the Puglia blue histogram at the week 40 in Figure 6. The estimation of the correct  infection rate is fundamental to understand in which stage the pandemic is. An \textit{artificial}  reduction in the infection rate not only creates the problem of not having the real picture of the infected in the territory but also creates another even more serious problem which is the \textit{artificial}  reduction of the pandemic reproduction factor, obtained by analysing the growth of the infection  rate.

Looking at Figure 6, the difference between white and blue histograms is evident for 5-Emilia Romagna, 6-Friuli Venezia Giulia, 9-Lombardia, 8-Liguria, 12-P.A. Bolzano, 13-P.A. Trento, 20-Veneto, and  21-Val d' Aosta: 8 of the 9 regions (with deaths per million greater than 1300) which appear in the top of the table given in Figure 2. Piemonte local authorities used, in addition to molecular tests, serological tests. This obviously reduces the infection rate. As previously observed such tests do not give information on the patient's current state of infection.  The number of serological tests was  removed later, see the Piemonte plot in  Figure 6 at the pandemic week number 43 (centred at December 17).  Once again an incomprehensible choice,  in this case done by the local authorities which did not follow the indication given by the national ones.  The difference between the white and blue histograms  is almost non-existent for 2-Basilicata, 3-Calabria, 4-Campania, 11-Molise, and 16-Sardegna,  very small for 17-Sicilia, and small for 7-Lazio and 15-Puglia. A minimal difference between the two curves clearly shows the local capacity  to control  the spreading of the pandemic.

Studying the pandemic reproduction factor is certainly one of the main objectives in facing an outbreak, but a correct analysis requires to know the number of new people tested by a molecular analysis. Looking at Figure 6, it is incomprehensible how Veneto was always placed in the area of low pandemic hazard zone (yellow) and Calabria placed, at November 4, in the one  of greatest pandemic danger (red) and then, at November 27, in the medium risk (orange)  never reached the yellow classification.  The infection rate is surely one of the first parameter to be investigated and it could play a fundamental role in  a timely anticipation of the territorial spreading  of the virus. Nevertheless, three others parameters are of great importance when facing the outbreak: Hospitalized per Million of Inhabitants (pMoI), hospitalized in ICUs pMoI, and, finally deaths pMoI.  The detailed study  of these 3 pandemic
parameters is the subject matter of the next section and leads, together with the confirmed over tested ratio, to the introduction of a numerical pandemic criticality index.

When discussing the relationship between new confirmed cases and new tested people, we used 5 colored  zones: up to 10\% (cyan), between 10\% and  20\% (green), between 20\% and 30\% (yellow), between 30\% and  40\% (orange), and, finally,  greater than 40\% (red), see Figure 5. Let us now determine the  criticality of the colored areas for hospitalized and  hospitalized in ICUs. For the hospitalized in ICUs, the  pandemic critical areas  have been determined by taking into account  the total number of ICU beds available in each region. Observing that part of the beds must be reserved for non CoVid-19 patients, the begin of the red zone was fixed at  60\% of  occupancy. The least criticality areas  are then found  between 45 \% and 60\% (orange), 30\% and 45\% (yellow), 15\% and 30\% (green), and, finally,  below 15\% (cyan).   In facing the second pandemic wave, regions have increased their capacity for ICU beds: 3-Calabria 152 ICUs (an increase of 
4.8\%), 9-Lombardia 983 (14.2\%), 15-Puglia 366 (20.4\%), and 21-Veneto 825 (67.0\%).
Considering their different populations (1.925 M, 10.104 M, 4.008 M, and 4.907 M) we thus have 79.0, 97.3, 91.3, and 168.1 ICUs per million of inhabitants. The colored zone of criticality
are then found at (11.9, 23.7, 31.6, and 47.4) for Calabria,   (14.6, 29.2, 39.9, and 58.4) for Lombardia,   (13.7, 27.4, 36.5, and 54.8) for Puglia, and  (25.2, 50.4, 67.2, and 100.9) for Veneto, see Figure 8.  

During the first pandemic wave the most affected region  showed a factor 10 of proportionality between hospitalized and hospitalized in ICUs (see for example the gray histograms
of 5-Emilia Romagna, 8-Liguria, 9-Lombardia, and 14-Piemonte in Figure 7 and 8). The criticality zones for hospitalized people were thus obtained from the ones of hospitalized in ICUs by using a  factor 10.

In Figure 9, we find the (7-day average) daily deaths for million of inhabitants with  colors bands fixed at 2, 4, 6, and 8. To control the pandemic curves the daily deaths should  not exceed 5 deaths per million of inhabitants. For one death, we approximatively find 50 hospitalized in ICUs and this mean an occupancy of 30\% for a country with 150 IUCs per million of inhabitants (the number suggested by the WHO).

\section{The pandemic criticality index}

The use of colors to graphically show the territorial pandemic criticality  is  understandable: 
Colored maps allow to immediately recognize which areas are in a critical situation. 
Nevertheless, the colored zones  should \textit{always} be accompanied  by their corresponding numerical pandemic criticality index. Going from one color to another is like making a quantum leap and it could even create confusion in the population. A number could better explain the evolution of the pandemic criticality in a given area.  In this Section, we will see how, by using the four parameters given in the previous Section,  it is possible to  introduce a numerical index of pandemic criticality.

Let us consider the following five criticality zones with the corresponding numerical index interval:\\

\begin{tabular}{crcl}
{\color{cyan} $\bullet$} &     low risk &-& [\,0\,,\,1\,)\,\,,\\
{\color{green} $\bullet$} &      medium  risk &-& [\,1\,,\,2\,)\,\,,\\
{\color{yellow} $\bullet$} &  medium/high risk &-& [\,2\,,\,3\,)\,\,,\\
{\color{orange} $\bullet$} &  high risk &-& [\,3\,,\,4\,)\,\,,\\
 {\color{red} $\bullet$}&   very high risk &-& [\,4\,,\,$\infty$\,)\,\,.\\
\end{tabular}
\vspace*{0.5cm}

\noindent The 7-day average infection rate bands are at 10\%, 20\%, 30\%, and 40\% (see Figure 6). So, the first normalised parameter to be used in calculating the numerical pandemic criticality is
\begin{equation}
 \rho_{\1}\,\,=\,\,\frac{1}{10\%}\,\,\frac{\textrm{new confirmed cases}}{\textrm{new tested people}}\,\,.
 \end{equation}
The ratio of 7-day average hospitalized  and hospitalized in ICUs  over the available ICUs in each region are characterized by the following critical values: 150\%, 300\%, 400\%, 600\% (Figure 7), and  15\%, 30\%, 40\%,  60\% (Figure 8).  In this case, we introduce the following two normalised parameters
\begin{eqnarray}
\rho_{\2}&=&\frac{1}{150\%}\,\,\frac{\textrm{hospitalized}}{\textrm{ICUs}}\,\,, \nonumber \\
 & & \\
 \rho_{\3}&=&\frac{1}{15\%}\,\,\frac{\textrm{hospitalized in ICUs}}{\textrm{ICUs}}\,\,.\nonumber 
\end{eqnarray}
Finally, the bands for the 7-day average daily deaths per million of inhabitants are found at 2, 4, 6, and 8. Consequently, the last normalised parameter is given by 
\begin{equation}
 \rho_{\4}\,\,=\,\,\frac{1}{2}\,\,\textrm{daily deaths per million}\,\,.
 \end{equation}
These 4 numerical parameters can be then weighed, leading to the numerical pandemic criticality index
 \begin{eqnarray}
\textrm{pci}_{a,b,c,d}& = & \frac{a\,\rho_{\1}+  b\,\rho_{\2}+ c\,\rho_{\3}+ d\,\rho_{\4}}{a+b+c+d} \,\,.
 \end{eqnarray}
In Figure 10, we plot the numerical index for (a,b,c,d)=(1,1,1,1) and (1,2,3,4), see white lines. In the same Figure, we also find the mean criticality index
 \begin{eqnarray}
\textrm{pci} & = & (\,\textrm{pci}_{1,1,1,1} +\, \textrm{pci}_{1,2,3,4}\,)\,/\,2 \,\,,
 \end{eqnarray}
see the blue dots in Figure 10. In Table 3, we give, for each region,  the pandemic criticality index corresponding to the last 15 pandemic weeks of 2020: From week 31 (centred at September 24) to week 45 (centred at December 31). In Figure 10 and Table 3, we also associate a gradation of colors to facilitate the graphical presentation of the pandemic criticality index.

By using the numerical pandemic criticality index, we can also check the containment measures adopted by the Italian government. At the beginning of November, the Italian authorities
 divided the national territory into 3 pandemic risk areas: Yellow, orange and red areas  with corresponding containment measures. Yellow area:  Curfew from 10pm to 5am of the following day, public transport with 50\% of occupancy, distance learning for high schools and face-to-face for middle and elementary schools, shopping centres closed on weekends (with the exception of pharmacies, tobacco shops, newspapers stands), bars and restaurants closing at 6pm. Orange area: to the containment measures of the yellow area we have to add  the closing of bars and restaurants and prohibition of travel between different municipalities. Red area:  prohibition of any type of movement if not justified. 

The first classification done by  the Italian authorities (November 4) determines the red zone for Piemonte, Lombardia, Val d'Aosta and Calabria, the orange one for Puglia and Sicilia, and the yellow area for all the other regions and the two autonomous provinces of Bolzano and Trento. Looking at Table 3, we  observe that, at week 36 (centred at October 29)  3-Calabria  had a numerical pandemic index of 0.53 (the best one in Italy) and 2 regions had an index greater than 3 (8-Liguria, 3.15, and 20-Val d'Aosta, 6.24). 
The Italian authorities determine the zone of Calabria to be red and the one of 
Liguria to be yellow. In the same Table, we also find that 21-Veneto, 1.34,  and 15-Puglia had practically the same pandemic index. One week later, Liguria turns into an orange zone. The new containment measures allowed to stop and the reduce its numerical pandemic index, see the pandemic week 39 (centred at November 19). For Veneto, always yellow zone, the pandemic index will increase during the following week reaching 6.14 at the last week of the year. Calabria never exceeds 2 (one of the best result in Italy) but it is one of the first regions to be classified as a red area and only at December 5 is turned into an orange zone. By comparing the pandemic criticality index given in  Table 3 with  the risk classification assigned by  the Italian authorities discussed in Section 3, the reader will find  many other inconsistencies.

\section{Conclusions}

The infection reproduction number (IRN) is only one of the parameter to be checked and probably the one most problematic. 
In the first wave of the pandemic the number of tested people worldwide was often not a \textit{reliable} number, this implied a wide and  sometimes embarrassing  IRN range.  During the second pandemic wave, the number of tests made has certainly become more reliable but as we have seen in this article, in Italy the tests made were used instead of the \textit{new tested people}.    It is clear that monitoring the territorial increase of the infection is essential to prevent the diffusion, but once the pandemic spreads on the territory other parameters are important to determine the appropriate containment measures. If a region, due to a precarious health system or to lack of heath professionals,   is, for example,  unable to treat patients, it is clear that, independently  of its infection reproduction number, containment measures must be taken. Hospitalized, hospitalized in ICUs, and, obviously, daily deaths  must not only be monitored systematically but also \textit{weighed} appropriately.  The use of a \textit{numerical} pandemic criticality index based on few and effective parameters can also be easily understood by population and media. Another important point is that, by using a numerical range instead of color bands, we can better differentiate  the criteria of social distancing. 
 \begin{center}
\includegraphics[width=8.5cm, height=11cm]{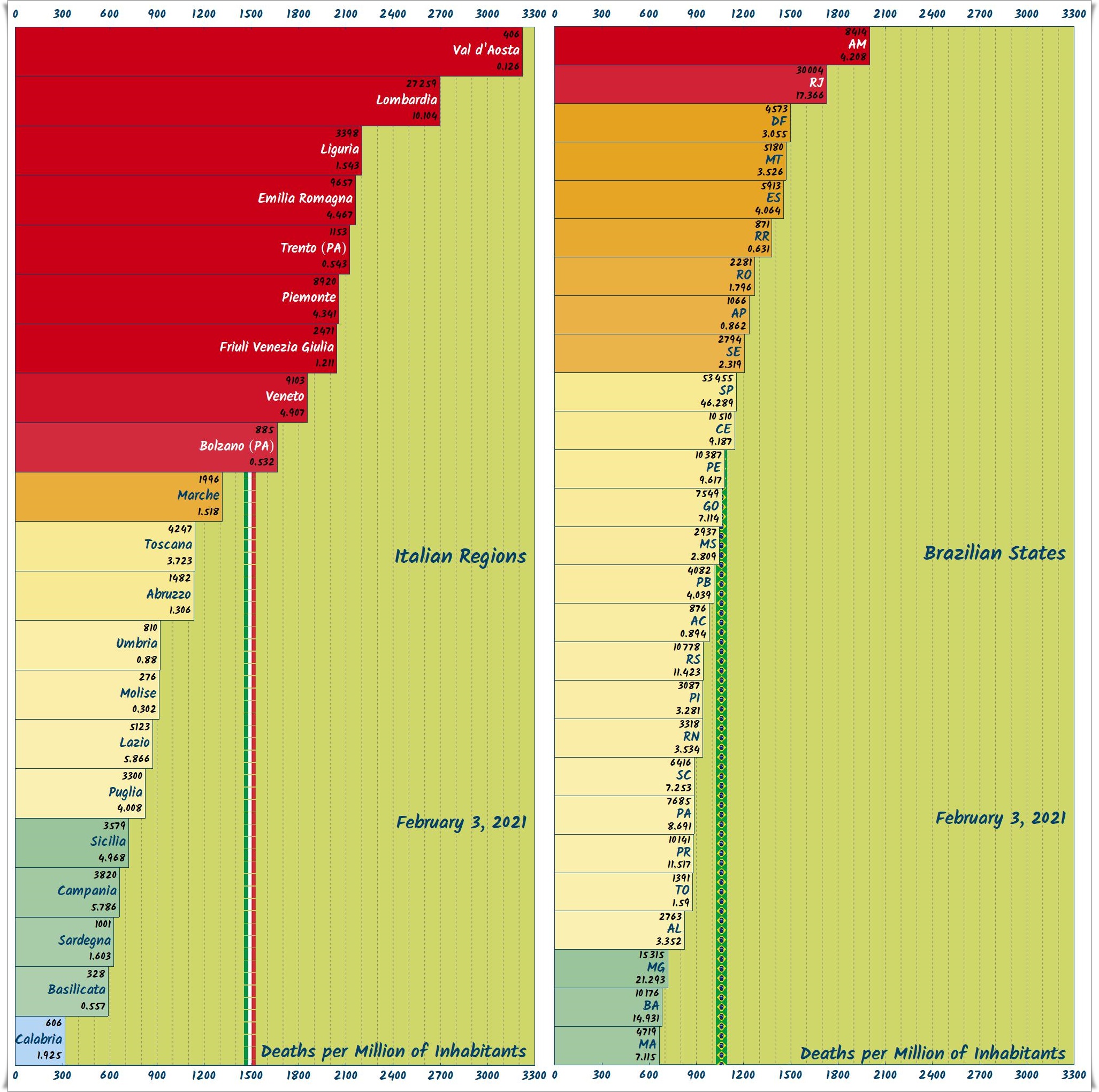}
\end{center}
CoViD19 web-page at Imecc/Unicamp (Prof. Stefano De Leo) \cite{sdl}.
 
\newpage

\subsection*{Acknowledgements}

The authors are deeply grateful to  Prof. Edmundo Capelas de Oliveira and Dr. Rita Katharina Kraus  for their scientific comments and suggestions  during the preparation of this article. The authors also thank Fapesp (SDL) and Capes (MPA) for the financial support.

\newpage

\section*{Appendix}
The list of 21 indicators for the pandemic risk control established by the ministerial decree of April 30\cite{MS}.

\begin{flushright}
\textit{(IA)}
\end{flushright}

1) Number of symptomatic cases notified per month for which the starting date of symptoms is indicated over the total  symptomatic cases notified to the surveillance system in the same period;

2) Number of cases notified per month with a history of hospital admission, in wards other than intensive care, for which  the date of admission is indicated  over  the total of cases with a history of hospital admission, in wards other than intensive care, notified to the to the surveillance system in the same period;

3) Number of cases notified per month with a history of transfer/admission to the ICU for which the date of transfer/admission is indicated over the   total of cases with a history of transfer/admission to the  ICU notified to the surveillance system;

4) Number of cases notified per month for  which the municipality of domicile or residence is reported over  the total number of cases notified to the surveillance system in the same period;

5) Number of checklists administered weekly to residential and health care facilities (optional);

6) Number of residential and health care facilities that respond to the checklist on a weekly basis with at least one criticality found (optional);

\begin{flushright}
\textit{(IB)}
\end{flushright}

7) Percentage of positive swabs, excluding as far as possible all screening activities and \textit{re-testing} of the same subjects, overall and by macro-setting (territorial, emergency room/hospital, other) per month.

8) Time between symptom onset date and diagnosis date.

9) Time between symptom onset date and isolation date (optional).

10) Number, type of professional figures and time  over the total number of people dedicated to contact-tracing in each territorial service.

11) Number, type of professional figures and time over the total of people dedicated in each territorial service to the activities of sampling/sending to the reference laboratories and monitoring of close contacts and cases placed respectively in quarantine and isolation.

12) Number of confirmed cases of infection in the Region for which a regular epidemiological investigation was carried out with the search for close contacts, out of the total of new confirmed cases of infection.

\begin{flushright}
\textit{(IC)}
\end{flushright}

13) Number of cases reported to the Civil Protection in the last 14 days.

14) $R_t$ calculated on the basis of integrated ISS surveillance (two indicators, based on the  symptom start  and hospitalization dates, are used).

15) Number of cases reported to Covid-net sentinel surveillance per week (optional).

16) Number of cases by diagnosis date and symptom onset date reported to integrated Covid surveillance per day.

17) Number of new transmission outbreaks (2 or more epidemiologically linked cases or an unexpected increase in the number of cases in a defined time and place).

18) Number of new cases of confirmed SARS-CoV-2 infection per region not associated with known transmission chains.

19) Number of accesses to the emergency room with ICD-9  classification compatible with syndromic pictures attributable to Covid-19 (optional)

20) Occupancy rate of total intensive care beds (code 49) for Covid patients.

21) Occupancy rate of total medical area beds for Covid patients.

\newpage

\FigW
\FigR
\TabA
\FigA
\TabBC
\FigB
\FigC
\FigD
\FigE
\FigF
\FigG
\TabCol
\FigH

\end{document}